\documentclass[aps,pra,twocolumn,footinbib]{revtex4-1}
\usepackage{graphicx}
\usepackage{indentfirst}
\usepackage{physics}
\usepackage{braket}
\usepackage{float}
\usepackage{amsmath}
\usepackage{epstopdf}
\usepackage{footnote}
\usepackage{CJK}
\usepackage{esint}
\usepackage{color}
\usepackage[T1]{fontenc}
\usepackage{subfigure}
\usepackage{amsfonts}
\usepackage{footmisc}
\usepackage{scrextend}
\usepackage{multirow}
\usepackage[hyperfootnotes=false]{hyperref}
\usepackage[acronym]{glossaries}

\newacronym{QN}{QN}{quantum network}
\newacronym{QI}{QI}{Quantum information}

\usepackage[english]{babel}
\usepackage{url}
\usepackage{bm}
\usepackage{hyperref}
\definecolor{darkblue}{rgb}{0,0,0.5}
\hypersetup{
colorlinks=true,
linkcolor=black,
filecolor=blue,
citecolor=darkblue,  
urlcolor=black,
}

\newcommand{\calC}{{\cal C}}

\newcommand{\calE}{{\cal E}}

\newcommand{\calN}{{\cal N}} 

\newcommand{\calG}{{\cal G}}

\newcommand{\1}{^{(1)}}

\def\be{\begin{equation}}
\def\ee{\end{equation}}
\def\ba{\begin{eqnarray}}
\def\ea{\end{eqnarray}}
\usepackage{bm}

\newcommand{\QZ}[1]{{{\textcolor{black}{#1}}}}
\newcommand{\QZR}[1]{{{\textcolor{black}{#1}}}}
\newcommand{\QZT}[1]{{{\textcolor{black}{#1}}}}

\begin{document}

\title{Quantum communication capacity transition of complex quantum networks}

\author{Quntao Zhuang$^{1,2}$}
\email{zhuangquntao@email.arizona.edu}
\author{Bingzhi Zhang$^{1,3}$}

\address{
$^1$Department of Electrical and Computer Engineering, University of Arizona, Tucson, Arizona 85721, USA
}
\address{
$^2$James C. Wyant College of Optical Sciences, University of Arizona, Tucson, AZ 85721, USA
}
\address{
$^3$Department of Physics, University of Arizona, Tucson, AZ 85721, USA
}

\begin{abstract}
Quantum network is the key to enable distributed quantum information processing. As the single-link communication rate decays exponentially with the distance, to enable reliable end-to-end quantum communication, the number of nodes needs to grow with the network scale. 
For highly connected networks, we identify a threshold transition in the capacity as the density of network nodes increases---below a critical density, the rate is almost zero, while above the threshold the rate increases linearly with the density. Surprisingly, above the threshold the typical communication capacity between two nodes is independent of the distance between them, due to multi-path routing enabled by the quantum network. In contrast, for less connected networks such as scale-free networks, the end-to-end capacity saturates to constants as the number of nodes increases, and always decays with the distance. Our results are based on capacity evaluations, therefore the minimum density requirement for an appreciable capacity applies to any general protocols of quantum networks. 
\end{abstract} 


\maketitle

\section{ Introduction}
\acrfull{QI} science has brought advantages in various applications~\cite{Shor_1997,Giovannetti2004,Bennett2002,gisin2007quantum}. 
To unleash the full power of \acrshort{QI} processing in distributed tasks~\cite{lynch1996distributed,andrews2000foundations}, a \acrfull{QN}~\cite{kimble2008quantum,biamonte2019complex,wehner2018quantum,kozlowski2019towards,miguel2020genuine} aiming at entanglement distribution and \acrshort{QI} transmission is the key.

The Internet is mainly built upon fiber networks, with photons as the information carrier. Similarly, photons as the only known ``flying qubits'' will likely be the information carrier in a \acrshort{QN}. In both cases, channel loss is the major challenge to communication. Therefore, networking protocols that make use of intermediate nodes or repeaters are important for both.
Unlike classical information, QI cannot be simply cloned and amplified, and therefore increasing the number of nodes, even repeater nodes~\cite{briegel1998quantum,jiang2009quantum,sangouard2011quantum,munro2015inside,muralidharan2016optimal,dias2017quantum,jiang2018,furrer2018repeaters,dias2020quantum,Seshadreesan2020,goodenough2020optimising}, is costly. In this regard, a key question for designing a \acrshort{QN} is to understand the trade-off between the density of nodes and the entanglement distribution rate: how many nodes are necessary to guarantee reliable \acrshort{QI} transmission between multiple users in a fixed region?

The answer not only depends on the overall distances between the users, but also on the topology of the \acrshort{QN} to be built~\cite{acin2007entanglement}. 
As it is likely that well-developed classical fiber networks can be adopted as the base of \acrshort{QN}s, Ref.~\cite{brito2020statistical} \QZ{developed a model for \acrshort{QN} on the the probabilistic transmission of single photons and took a classical network science approach to study its connectivity by the giant component.} However, for \acrshort{QN}s exploiting quantum technologies such as quantum error correction~\cite{calderbank1996} and non-classical state generation~\cite{caves1981quantum,gottesman2001encoding}, the semi-classical approach \QZT{has a limited implication}. \QZT{In particular, Ref.~\cite{brito2020statistical}'s critical density highly depends on the number of repetitions of each channel use and thus blurs the essential constraints.} More recently, Refs.~\cite{rabbie2020designing,coutinho2021} considered effects from repeater nodes. As the results rely on specific protocols, the fundamental limits of the trade-off remains unclear.
We address the same question with a full QI approach based on the fundamental limits~\cite{pirandola2009,takeoka2014fundamental,pirandola2017fundamental,pirandola2019end,azuma2016fundamental}, \QZT{and obtain a minimum density requirement that generally applies to any protocols. }

As the exact architecture and protocols of \acrshort{QN}s are unclear, we take the information-theoretical approach and evaluate the end-to-end capacity~\cite{pirandola2019end} of \acrshort{QI} transmission. To account for different possibilities of the future \acrshort{QN}s, we consider typical types of network models~\cite{pastor2007evolution}, based on the Waxman networks~\cite{waxman1988routing,lakhina2003geographic}, \QZ{Erd\H{o}s-R\'enyi model} and scale-free networks~\cite{barabasi1999emergence,yook2002modeling}. 
Our results \QZ{provide an upper bound to characterize the quantum capacity of \acrshort{QN}s and the analysis applies} to all kinds of quantum communication.
In Waxman and \QZ{Erd\H{o}s-R\'enyi} \acrshort{QN}s, the ensemble-averaged capacity abruptly transits from almost zero to nonzero values at a critical density of nodes. Above the threshold, it grows with the density linearly, at a rate depending on the statistical properties of the \acrshort{QN}. Surprisingly, in this region the end-to-end capacity typically does not depend on the distance between the two end nodes, due to the multi-path routing enabled by the coordination of the entire \acrshort{QN}. In scale-free \acrshort{QN}s, the ensemble-averaged capacity saturates to a constant depending on the scale of the network as the density of nodes increases, due to the limited connectivity in the network that prevents efficient multi-path routing.

\section{Model of \acrshort{QN}s} The skeleton of a \acrshort{QN} can be described by a graph, with vertices $\calG$ being the network nodes and edges $\calE$ representing the transmission links~\cite{zhang2020entanglement}. As nodes are located geographically, we can assign a 2-D coordinate $\bm x$ to each node. The transmission link along each edge $E_{\bm x, \bm x^\prime}$ is modeled as a bosonic pure loss channel, with a transmissivity $\eta(\bm x, \bm x^\prime)=10^{-\gamma D(\bm x, \bm x^\prime)}$ for fiber length $D(\bm x, \bm x^\prime)$ at a state-of-the-art rate $\gamma=0.02$ per kilometer (km). For simplicity, we assume that for each edge, the fiber length $D(\bm x, \bm x^\prime)$ and the geographical distance $\|\bm x-\bm x^\prime\|_2$ are identical.

With the transmission links on each edge defined, one needs to specify the graph structure---the coordinates and connections of the vertices---to specify the \acrshort{QN}. Without loss of generality, we choose the coordinates $\bm x$ of the $N$ nodes uniformly random in a square $\Omega_R\equiv [-R,R]\times [-R,R]$, with an area of $|\Omega_R|=4R^2$.

In the random Waxman model~\cite{waxman1988routing,lakhina2003geographic}, each pair of nodes is connected with a probability $\Pi\left(\bm x, \bm x^\prime \right)= e^{-D(\bm x, \bm x^\prime)/\alpha L}$ decaying exponentially with the distance. Here $L=2\sqrt{2}R$ is the maximum possible distance in a square; the constant $\alpha$ controls the typical fiber length and is fixed so that $\alpha L=226$km to model the U.S. fiber-optics networks~\cite{lakhina2003geographic}. \QZT{It is worthy to point out that Ref.~\cite{brito2020statistical} adopted the same Waxman \acrshort{QN}s.}
In the scale-free model~\cite{yook2002modeling}, the network is built up dynamically: when each node $\bm x$ is being added, it is connected to $m$ nodes out of all the previous added nodes. The probability of node $\bm x^\prime$ being connected to node $\bm x$ is proportional to the current degree $D_g\left(\bm x^\prime\right)$ and inversely proportional to the distance $D\left(\bm x, \bm x^\prime\right)$, i.e.,
$ 
\Pi\left(\bm x, \bm x^\prime \right)\propto D_g\left(\bm x\right)/D\left(\bm x, \bm x^\prime\right),
$ 
in contrast to the Waxman model's exponential decay with distance.

\begin{figure}
    \centering
    \includegraphics[width=0.5\textwidth]{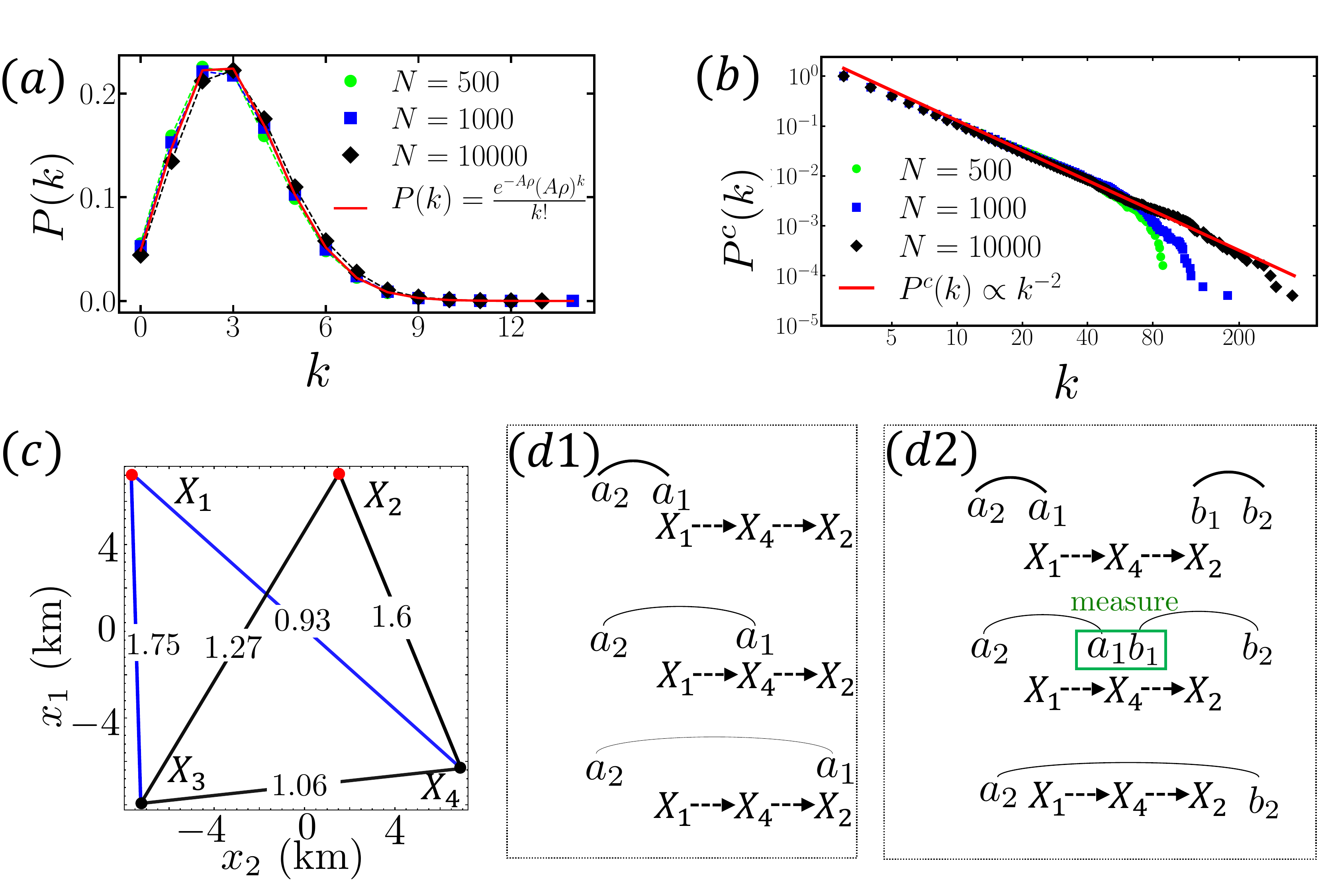}
    \caption{(a) The degree distribution of Waxman network, with density $\rho=10^{-5}$, 
    fits well with a Poisson distribution (red curve) with $A=3.0\times 10^5$.
    (b) 
    The cumulative degree distribution of the scale-free model, with density $\rho=10^{-5}$, fits well with a power-law (red curve).
    (c) A four-node \acrshort{QN}, with the axes as the geographical coordinates. The blue color indicates a cut between \QZ{$X_1$} and \QZ{$X_2$} and the number on each edge equals the edge capacity in Eq.~\eqref{C_edge}. For the cut indicated by blue edges, the cut capacity of Eq.~\eqref{C_end_to_end} $\calC\left(\QZ{\mathbb{U}}_{\bm x, \bm x^\prime}\right)=1.75+0.93=2.68$, which turns out to be the minimum cut.
    (d1-d2) Examples of entanglement distribution protocols. (d1) Direct communication strategy with potential error correction.
    (d2) Entanglement-swap strategy. After an entanglement swap measurement on $a_1b_1$, nodes $X_1$ and $X_2
    $ can share an entangled state in $a_2b_2$. 
    \label{fig:example}
    }
\end{figure}

\begin{figure*}
\centering
\includegraphics[width=1\textwidth]{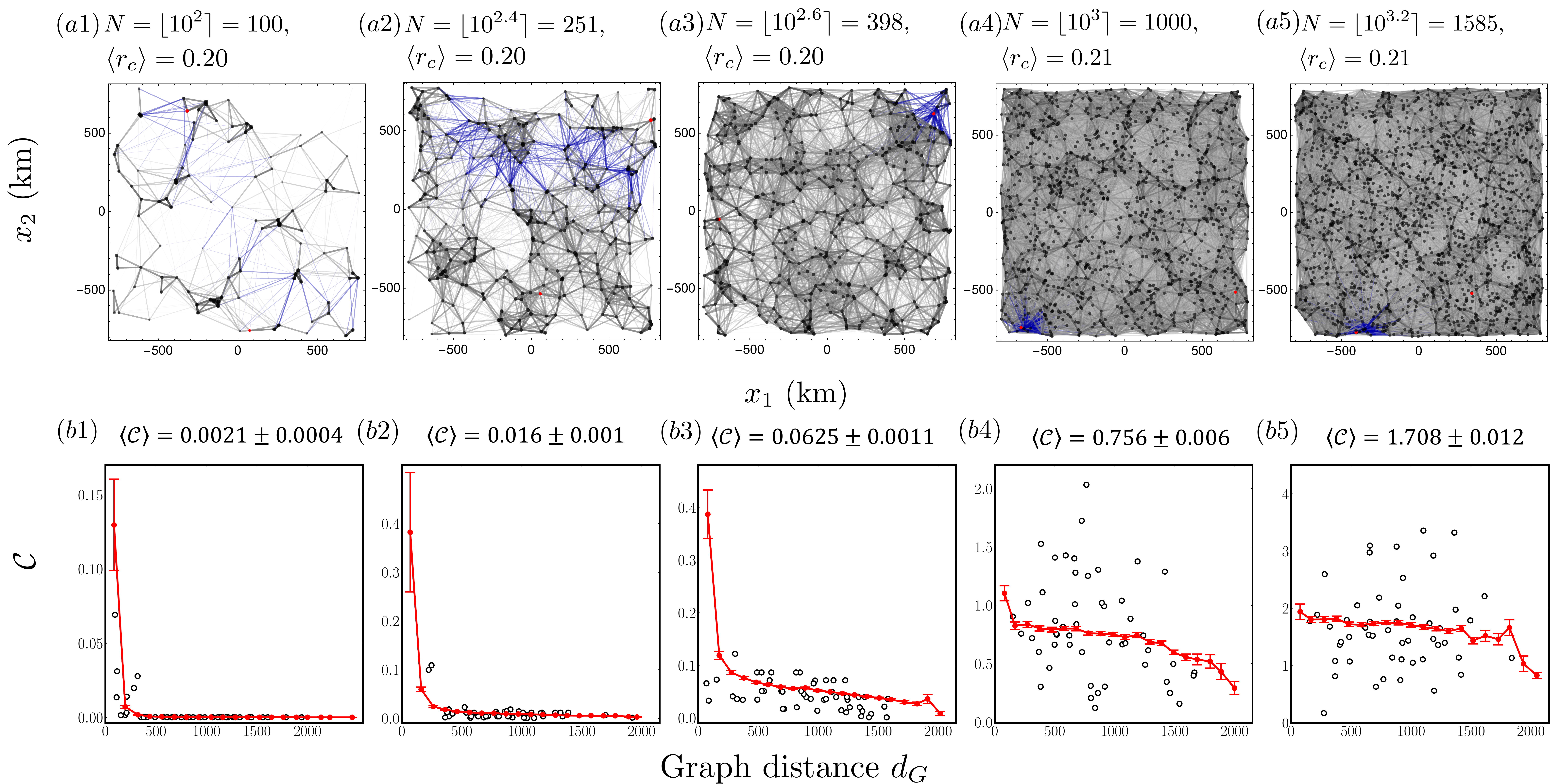}
\caption{Waxman \acrshort{QN}s, $\alpha=0.1$ ($R\simeq 800$ km). (a1)-(a5) Visualizations with different number of nodes $N$. The darkness and opacity of the color of the nodes and edges indicate the relative amplitude of the capacity (darker means larger).
The blue edges are the minimum cut solution to random pairs of end nodes indicated by the red dots.
(b1)-(b5) The end-to-end capacity $\calC\left(\bm x, \bm x^\prime\right)$ between random nodes $\bm x, \bm x^\prime$ for \acrshort{QN}s with fixed $N,\alpha$. The x-axis is the graph distance between two nodes $d_G\left(\bm x,\bm x^\prime\right)$, in terms of the shortest fiber path. The scattered circles are 50 random pairs in a single \acrshort{QN} sample and the red dashed lines indicate the average obtained from 5000 random data. 
\label{fig:graphs}
\label{fig:graphs_full}
}
\end{figure*}

To obtain a direct impression, we visualize the two models in Fig.~\ref{fig:graphs} (a) and Fig.~\ref{fig:graphs_YOOK} (a) respectively. Immediate differences in the connectivity can be seen, e.g. by comparing Fig.~\ref{fig:graphs} (a2) and Fig.~\ref{fig:graphs_YOOK} (a1): for the same $N=1585$ nodes in a region of scale $R\simeq 800$ km, the Waxman model is much more connected and homogeneous, while the scale-free model is less connected and heterogeneous. These differences can be captured by their statistical properties. As shown in Fig.~\ref{fig:example}(a)(b), the Waxman \acrshort{QN} model has a Poisson degree distribution and the average degree grows with the number of nodes linearly~\cite{poisson}; while the scale-free \acrshort{QN} model has a long-tailed power-law degree distribution and a bounded average of $2m$. It is also worthy mentioning that the Waxman model has a percolation phase transition~(see Appendix~\ref{appA}), where the percentage of the giant component of the graph increases sharply from close to zero to unity as the density $\rho=N/|\Omega_R|$ increases above a critical value of $\rho_G\simeq 7\times 10^{-6}$. However, we show that this necessary condition is far from being sufficient.

While we base our \acrshort{QN} models on the Internet, a \acrshort{QN} will be majorly different from Internet. In particular, classical repeaters~\cite{mukherjee2000wdm} are not counted as network nodes in the study of Internet~\cite{pastor2007evolution}, as they are universally deployed and cheap. In contrast, quantum repeaters are nontrivial and therefore directly considered as network nodes in this study. In this regard, the Waxman model's exponential decay of long direct links will be more likely for \acrshort{QN}s. However, our goal is {\em not} to determine which model can better represent a \acrshort{QN}, an emerging technology, but to characterize each model in terms of quantum communications.

\section{Protocols and capacity formula}
To distribute entanglement between two nodes \QZ{$X_1$ and $X_2$} in a \acrshort{QN}, the nodes can transmit quantum states between all links and perform two-way classical communication in combination of local operations at each node. To begin with, let's consider an instance of a four-node network in Fig.~\ref{fig:example}(c).
In a single-path routing strategy, one can choose a path from \QZ{$X_1$ to $X_2$ (e.g. $X_1-X_4-X_2$, $X_1-X_3-X_2$, or $X_1-X_3-X_4-X_2$)} and utilize all the channels along the path once to distribute the entanglement. With the path fixed, one can either perform direct communication or adopt entanglement swap~\cite{zukowski1993event}, as shown in Fig.~\ref{fig:example}(d).
A more efficient approach is to adopt multi-path routing. For example, nodes \QZ{$X_1$ and $X_2$} in Fig.~\ref{fig:example}(c) can utilize multiple non-overlapping paths simultaneously \QZ{($X_1-X_3-X_2$ and $X_1-X_4-X_2$)} and achieve a better performance.

As protocols vary, to obtain universal results, we consider the ultimate achievable entanglement distribution rate among {\em all} protocols~\cite{pirandola2017fundamental,pirandola2019end}. In contrast to classical communication~\cite{GiovannettiV2014,Giovannetti2004}, \acrshort{QI} transmission rate for each edge is fundamentally limited by the channel loss to be
\be 
\QZ{\calC_E} \left(E_{\bm x, \bm x^\prime}\right)=-\log_2\left(1-\eta\right)=
-\log_2\left(1-10^{-\gamma D(\bm x, \bm x^\prime)} \right),
\label{C_edge}
\ee 
regardless of the energy, where $\eta=10^{-\gamma D(\bm x, \bm x^\prime)}$ is the channel loss~\cite{pirandola2017fundamental}. 
To characterize the importance of a single node, we define the node capacity $\QZ{\calC_N}\left(\bm x\right)=\sum_{\bm x^\prime \in \calN(\bm x)} \QZ{\calC_E}\left(E_{\bm x, \bm x^\prime}\right)$, as the sum of the edge capacities.

Consider the graph with edge capacities $\{\QZ{\calC_E} \left(E_{\bm x, \bm x^\prime}\right)\}$ as the weights (e.g. Fig.~\ref{fig:example}(c)), the problem of solving the end-to-end capacity is reduced to solving the minimum cut~\cite{pirandola2019end}.
Let's first introduce a cut $\QZ{\mathbb{U}}_{\bm x,\bm x^\prime}$ between two nodes $\bm x$ and $\bm x^\prime$ as the set of edges such that their deletion will disconnect the two nodes. For example, in Fig.~\ref{fig:example}(c), the blue part indicates a cut for $A$ and $B$. Then the capacity between end nodes $\bm x$ and $\bm x^\prime$ is given by the ``edge connectivity'' between them~\cite{pirandola2019end}
\be 
\calC \left(\bm x, \bm x^\prime\right)=
\min_{\QZ{\mathbb{U}}_{\bm x, \bm x^\prime}} \calC_U\left(\QZ{\mathbb{U}}_{\bm x, \bm x^\prime}\right)
\equiv 
\min_{\mathbb{U}_{\bm x, \bm x^\prime}}
\sum_{E_{\bm y, \bm y^\prime}\in \mathbb{U}_{\bm x, \bm x^\prime} } \calC_E \left(E_{\bm y, \bm y^\prime}\right).
\label{C_end_to_end}
\ee 
To obtain further insights, we derive an upper bound of the end-to-end capacity by the node capacities of the two end nodes,
$
\calC \left(\bm x, \bm x^\prime\right)\le  \min\left\{\QZ{\calC_N}\left(\bm x\right),\QZ{\calC_N}\left(\bm x^\prime\right)\right\},
$
as one can always choose the cut that consists of all edges connected to one of the end nodes.

\begin{figure}
\centering
\includegraphics[width=0.5\textwidth]{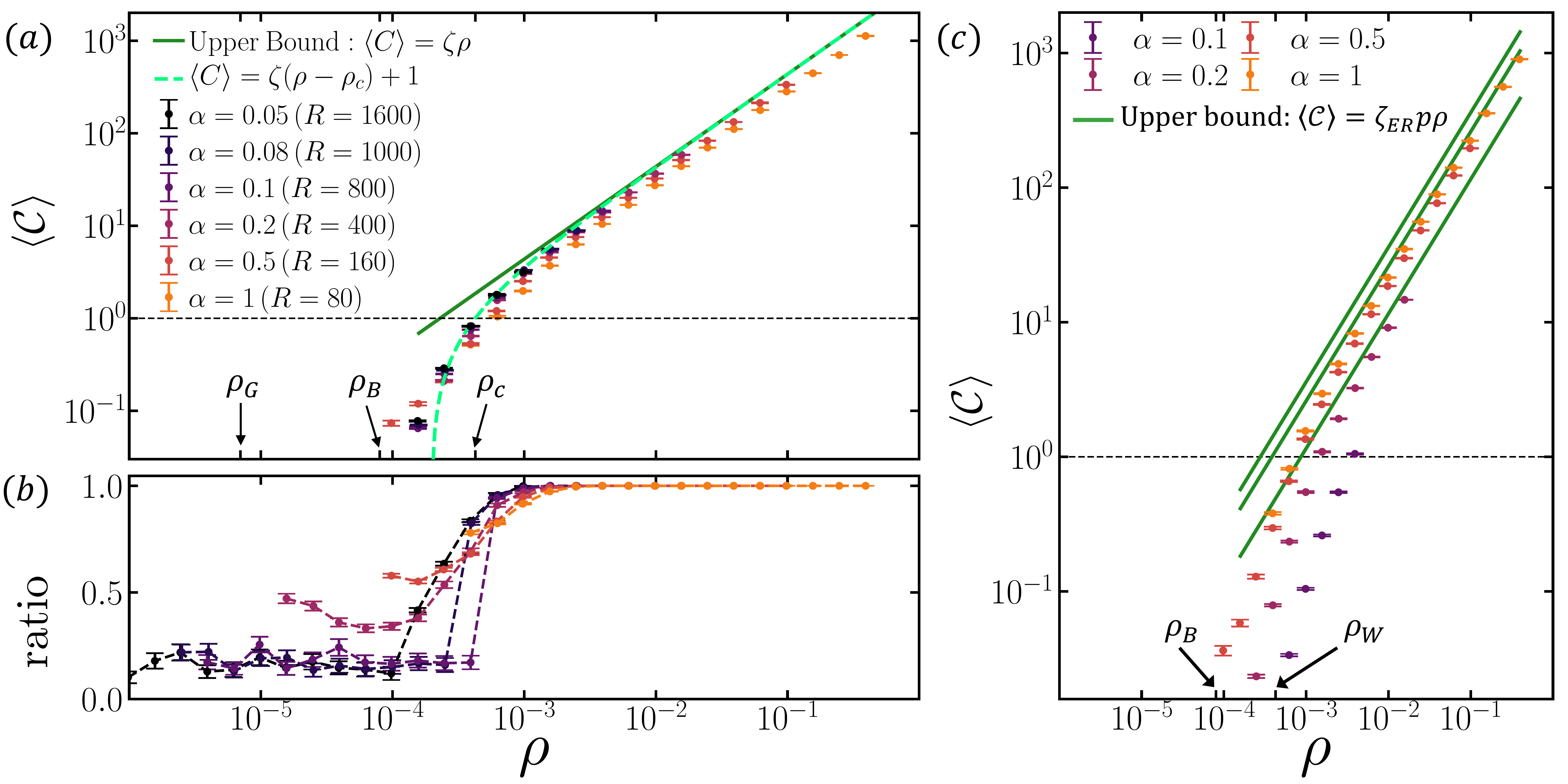}
\caption{(a) Average end-to-end capacity $\braket{\calC}$ vs. nodes density $\rho$. The dark green solid line and the light green dashed line give the upper bounds $\braket{\calC\left(\bm x\right)}\simeq \zeta \rho$ and its shifted fitting $\braket{\calC}=\zeta(\rho-\rho_c)+1$ respectively. The arrows indicate critical densities for the birth of giant connected component ($\rho_G$), for the prediction of Ref.~\cite{brito2020statistical} ($\rho_B$) and $\rho_c\simeq 4.25\times10^{-4}$ is when $\braket{\calC}=1$.
(b) The average of the ratio of end-node edges inside the minimum cut. It shares the same legend as in (a).
\QZ{(c) Average end-to-end capacity $\braket{\calC}$ of Erd\H{o}s-R\'enyi model vs node density $\rho$. The green lines from top to bottom correspond to the asymptotic upper bound $\braket{\calC} = \zeta_{ER}p\rho$ for $\alpha = 1, 0.5, 0.2$.}
\label{fig:C_rho_alpha}
}
\end{figure}

We take a statistical approach and evaluate the average end-to-end capacity $\braket{\calC \left(\bm x, \bm x^\prime\right)}$ in an ensemble of network models, where the average is over the choices of the end nodes $\bm x$, $\bm x^\prime$ and the random realization of the network, with fixed numbers of nodes $N$ and scale $R$. In this regard,
\be 
\braket{\calC \left(\bm x, \bm x^\prime\right)}\le  \braket{\min\left\{\QZ{\calC_N}\left(\bm x\right),\QZ{\calC_N}\left(\bm x^\prime\right)\right\}}\le \braket{\QZ{\calC_N}\left(\bm x\right)},
\label{C_end_to_end_UB_further}
\ee 
upper bounded by the ensemble-averaged node capacity. \QZT{Compared to the edge connectivity approach based on probabilistic single-photon transmission in Ref.~\cite{brito2020statistical}, our quantum capacity approach applies to all protocols and reveals essential features of a network.}

\section{Rate transition of Waxman \acrshort{QN}s} 
To study Waxman \acrshort{QN}s, we first fix the scale $R\simeq 800$ km and vary the number of nodes $N$. In Fig.~\ref{fig:graphs} (b), we plot the end-to-end capacity $\calC \left(\bm x, \bm x^\prime\right)$ of random pairs vs. the graph distance $d_G\left(\bm x,\bm x^\prime\right)$ (the shortest path length) between them. When the number of nodes is small (e.g. Fig.~\ref{fig:graphs} (b1)), the capacity decays with the graph distance drastically; while surprisingly, when the number of nodes becomes larger (e.g. Fig.~\ref{fig:graphs} (b2)), the capacity is almost independent of the graph distance~\cite{patil2020}. This is due to the effect of multi-path routing---the number of possible paths increases significantly with distance when the nodes are dense.

\begin{figure}
\centering
\includegraphics[width=0.5\textwidth]{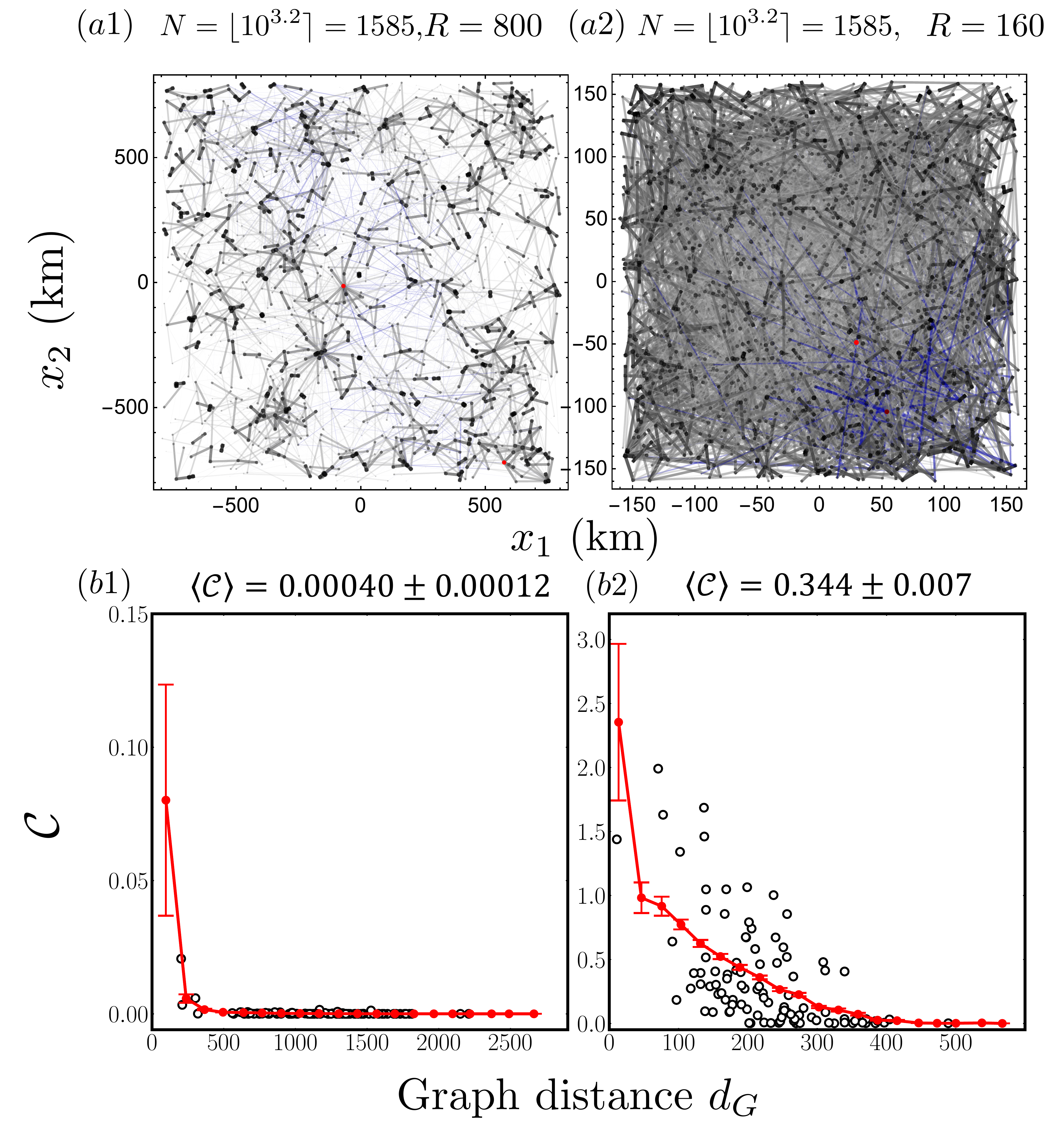}
\caption{Scale-free \acrshort{QN}s with $N=1585$ nodes, with similar arrangements and ensemble-averaging to Fig.~\ref{fig:graphs}. (a1)-(a2) Visualizations of the \acrshort{QN} model with different scales $R=800,160$ km, with fixed $N=1585$ nodes. 
(b1)-(b2) The corresponding end-to-end capacity. 
\label{fig:graphs_YOOK}
}
\end{figure}

To systematically evaluate the transition in the end-to-end capacity, we evaluate the ensemble-averaged capacity $\braket{\calC \left(\bm x, \bm x^\prime\right)}$ for different values of $R$ and $N$. We expect the density of nodes $\rho$ to be the crucial parameter. Indeed, we can show that when $R$ is large, the ensemble-averaged node capacity $\braket{\QZ{\calC_N}\left(\bm x\right)}\simeq \zeta \rho$, as the upper bound in Ineq.~\eqref{C_end_to_end_UB_further}, is linear in density $\rho$ with the coefficient $\zeta\simeq 4358$~(see Appendix~\ref{appC}).

In Fig.~\ref{fig:C_rho_alpha} (a), we plot the average capacity vs the node density $\rho$ for different system size $R$. Overall, for a fixed density $\rho$, the capacity $\braket{C}$ converges as the scale $R$ increases. When the density is small, the capacity is mostly close to zero~(see Appendix~\ref{appB}); As the density increases, we see a sudden transition from almost zero capacity to $o(1)$ capacity at a critical density. \QZR{The transition happens at around $\braket{C}\sim 1$ corresponding to a density} $\rho_c\simeq 4.25\times 10^{-4}$, which is much larger than the giant component transition $\rho_G\simeq 7 \times 10^{-6}$ and the result $\rho_B\simeq 6.82\times 10^{-5}$ from Ref.~\cite{brito2020statistical}.

After this transition, the average capacity increases linearly with node density $\rho$, approaching the upper bound $\zeta \rho$ (dark green line). The reason of the convergence can be observed from Fig.~\ref{fig:graphs} (a): when the connectivity is high, the minimum cut (blue edges) becomes a cut formed by all the edges connecting to one of the end points. To be more quantitative, we calculate the ratio of the edges in the minimum cut that contain at least one end node. As shown in Fig.~\ref{fig:C_rho_alpha} (b), the ratio transits from close to zero to unity at the same time as the end-to-end capacities approach the upper bounds. In fact, we find that a shifted upper bound $\zeta(\rho-\rho_c)+1$ fits the overall numerical results well, as shown by the green dashed line in Fig.~\ref{fig:C_rho_alpha}(a).

\QZT{Note that Ref.~\cite{brito2020statistical}'s critical density depends on the protocol parameters---e.g. the number of repetition $n_p$ for each link; therefore the value of their critical density is not an essential characterization of the \acrshort{QN}. Their results have to obey the constraint in our paper, as any protocol has its rate bounded by the capacity. We can confirm as follows: as they consider $n_p=1000$ repeated use of each channel to successfully establish one single Bell pair, the end-to-end capacity per channel use in their protocol is merely $10^{-3}$ for density $\rho=\rho_B$, which is in fact within the vanishing capacity region in our results. }





\section{Rate saturation of scale-free \acrshort{QN}s} 
\begin{figure}
\centering
\includegraphics[width=0.5\textwidth]{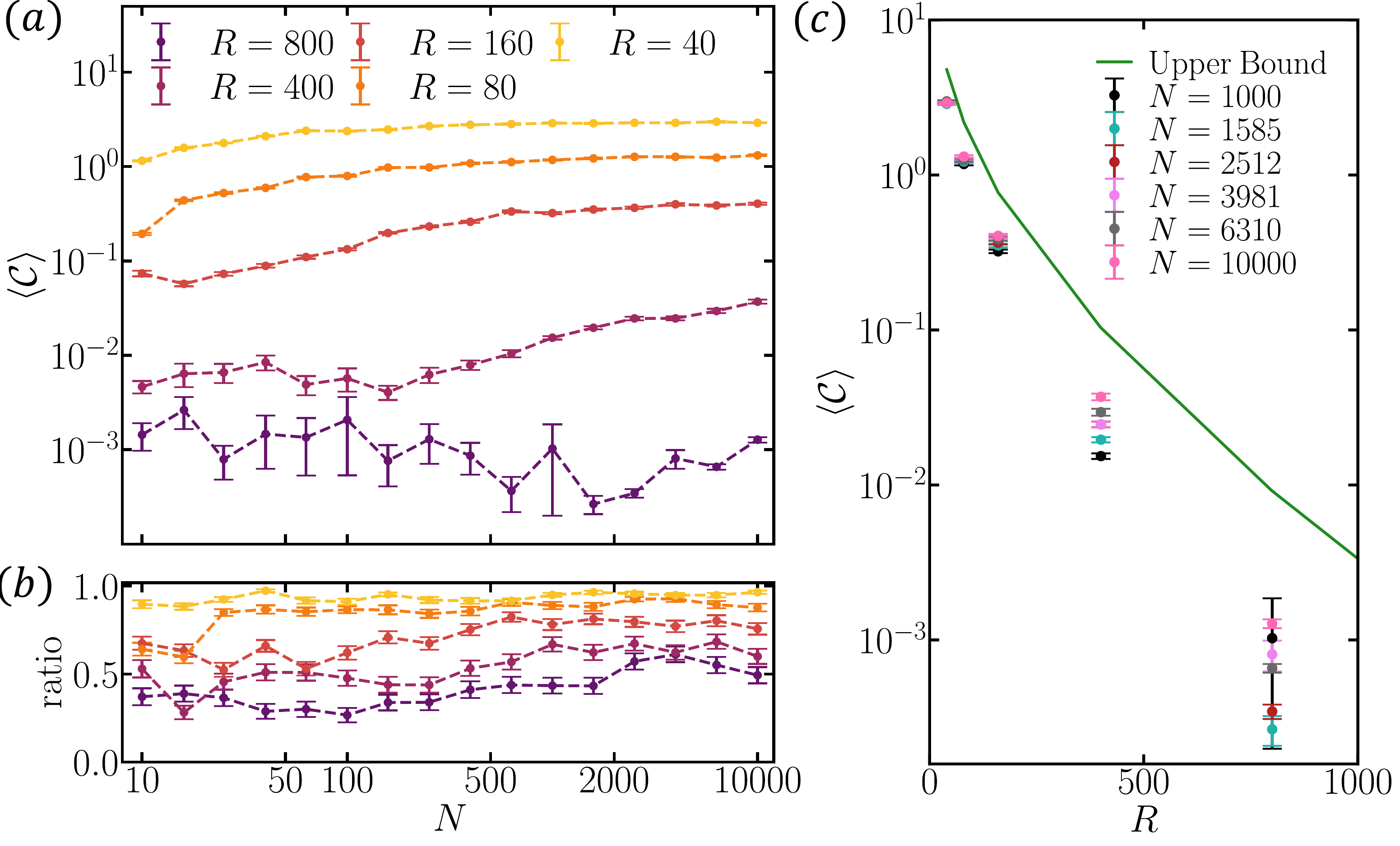}
\caption{Scale-free model. (a) Average end-to-end capacity $\braket{\calC}$ vs. the number of nodes $N$ for various scales $R$'s. 
(b) The average of the ratio of end-node edges inside the minimum cut.
(c) Capacity vs. the scale of the \acrshort{QN}. The orange curve is the upper bound $\braket{\calC\left(\bm x\right)}$ from numerical integration~(see Appendix~\ref{appE}). 
\label{fig:C_N_R_YOOK}
}
\end{figure}

Now we switch the focus to scale-free \acrshort{QN}s (see Fig.~\ref{fig:graphs_YOOK}). Similarly, we evaluate the end-to-end capacity for the same set of choices of $R$ and $N$. In Fig.~\ref{fig:C_N_R_YOOK} (a), the ensemble-averaged capacity $\braket{\calC(\bm x, \bm x^\prime)}$ grows as $N$ increases and saturates to a constant dependent on the scale $R$ of the network. This is due to the limited degree of scale-free networks, which constrains the upper bound of the node capacity to be bounded by a constant $\propto m$ and dependent on $R$~(see Appendix~\ref{appE}). As we can see in Fig.~\ref{fig:C_N_R_YOOK} (b), the ratio of edges of end points being in the minimum cut is now determined by the network scale, and gets close to unity when the network is small. Indeed, in Fig.~\ref{fig:C_N_R_YOOK} (c) we see the gap between the saturated capacity and the upper bound from node capacity is small for small $R$, while larger with $R$ increasing. Overall, the capacity decays with $R$ exponentially, even when the number of nodes is large.

In additional to the saturation of capacity, the graph-distance-independence of the capacity is absent for scale-free \acrshort{QN}s. In Fig.~\ref{fig:graphs_YOOK} (b), regardless of the capacity being large or small, there is a sharp decrease of the end-to-end capacity as the graph distance increases, in contrast to the Waxman \acrshort{QN}s in Fig.~\ref{fig:graphs} (b). This is due to the lack of multi-path routing, constrained by the connectivity of the scale-free networks. Indeed, we can find the average clustering coefficient $\braket{r_c}$ decaying with the system size, instead of saturating to constants with number of nodes in the Waxman case~(see Appendix~\ref{appA}).

 
\section{ Rate transition of Erd\H{o}s-R\'enyi \acrshort{QN}s}
We also extend our analyses to the Erd\H{o}s-R\'enyi model, a network model with uniform edge connection probability $p$. To compare with the Waxman model, we match the number of edges in Erd\H{o}s-R\'enyi model to the Waxman model with same $\alpha$ and $N$, via choosing a proper $p$. The corresponding degree distribution is binomial~(see Appendix~\ref{appA}).

We evaluate the transition of average end-to-end capacity $\braket{\calC}$ with node density in Fig.~\ref{fig:C_rho_alpha}(c), and identify similar trend to the Waxman model: when $\rho$ is large, $\braket{\calC}$ grows linearly with $\rho$; while when $\rho$ is small, there is still a sharp decrease in the capacity. While in the Waxman model, the capacity $\braket{\calC}$ agrees among different $\alpha$ in the linear transition, Erd\H{o}s-R\'enyi model shows a clear dependence on $\alpha$, and thus on the connection probability $p$. We can also explore further the dependence on $\alpha$ through the upper bound of node capacity in Eq.~\eqref{C_end_to_end_UB_further}.
$
\braket{\calC_N\left(\bm x\right)}\simeq \zeta_{ER} p \rho,
$
where $\zeta_{ER} \simeq 5137.9$ (see Appendix~\ref{appD}). We can directly see the dependence of $\braket{\calC}$ on connection probability $p$ from the asymptotic upper bound and we show them in Fig.~\ref{fig:C_rho_alpha}(c).

\section{Conclusion and discussions} 
In this paper, we examine the end-to-end quantum communication capacity in Waxman, \QZ{Erd\H{o}s-R\'enyi} \acrshort{QN}s and scale-free \acrshort{QN}s. Our results provide guidance on the design of \acrshort{QN} infrastructure, as the capacity places an achievable upper bound on rates of quantum communication protocols. 

In particular, our results suggest that when the connectivity of the \acrshort{QN} is high (like in the Waxman case), multi-path routing will enable reliable quantum communication. On the practical side, considering that quantum repeaters might be as costly and expensive as user nodes, this indicates that at a moderate metropolitan scale where users are dense and direct links are possible, it might be better to simply build more direct links between the users and utilize the multi-path routing for reliable quantum communication. 

\QZT{Our results are based on network capacity results and therefore reveals essential property of a \acrshort{QN}, independent on the protocol. We reveal more detailed properties of \acrshort{QN}s, other than the simple connectivity properties in Ref.~\cite{brito2020statistical}.}
\QZR{Our results address the entanglement generation capacity, which is the most relevant quantity in a \acrshort{QN}. In particular, our results allow unlimited two-way classical communication (via an underlying classical network) as assistance in the entanglement generation process. Ref.~\cite{brito2020statistical} limits the protocols to be at a single photon level, and is strongly dependent on the specific protocol parameters to generate entanglement. The density of nodes to guarantee reliable communication would depend on the exact meaning of reliable communication, however, a network above the threshold we identified is preferable as the capacity starts to become distance-independent.}

\begin{acknowledgments}
This research is supported by Defense Advanced Research Projects Agency (DARPA) under Young Faculty Award (YFA) Grant No. N660012014029, National Science Foundation (NSF) Engineering Research Center
for Quantum Networks Grant No. 1941583. Q.Z. also acknowledges Craig M. Berge Dean's Faculty Fellowship of University of Arizona. 
\end{acknowledgments}

\begin{thebibliography}{43}%
\makeatletter
\providecommand \@ifxundefined [1]{%
 \@ifx{#1\undefined}
}%
\providecommand \@ifnum [1]{%
 \ifnum #1\expandafter \@firstoftwo
 \else \expandafter \@secondoftwo
 \fi
}%
\providecommand \@ifx [1]{%
 \ifx #1\expandafter \@firstoftwo
 \else \expandafter \@secondoftwo
 \fi
}%
\providecommand \natexlab [1]{#1}%
\providecommand \enquote  [1]{``#1''}%
\providecommand \bibnamefont  [1]{#1}%
\providecommand \bibfnamefont [1]{#1}%
\providecommand \citenamefont [1]{#1}%
\providecommand \href@noop [0]{\@secondoftwo}%
\providecommand \href [0]{\begingroup \@sanitize@url \@href}%
\providecommand \@href[1]{\@@startlink{#1}\@@href}%
\providecommand \@@href[1]{\endgroup#1\@@endlink}%
\providecommand \@sanitize@url [0]{\catcode `\\12\catcode `\$12\catcode
  `\&12\catcode `\#12\catcode `\^12\catcode `\_12\catcode `\%12\relax}%
\providecommand \@@startlink[1]{}%
\providecommand \@@endlink[0]{}%
\providecommand \url  [0]{\begingroup\@sanitize@url \@url }%
\providecommand \@url [1]{\endgroup\@href {#1}{\urlprefix }}%
\providecommand \urlprefix  [0]{URL }%
\providecommand \Eprint [0]{\href }%
\providecommand \doibase [0]{https://doi.org/}%
\providecommand \selectlanguage [0]{\@gobble}%
\providecommand \bibinfo  [0]{\@secondoftwo}%
\providecommand \bibfield  [0]{\@secondoftwo}%
\providecommand \translation [1]{[#1]}%
\providecommand \BibitemOpen [0]{}%
\providecommand \bibitemStop [0]{}%
\providecommand \bibitemNoStop [0]{.\EOS\space}%
\providecommand \EOS [0]{\spacefactor3000\relax}%
\providecommand \BibitemShut  [1]{\csname bibitem#1\endcsname}%
\let\auto@bib@innerbib\@empty
\bibitem [{\citenamefont {Shor}(1997)}]{Shor_1997}%
  \BibitemOpen
  \bibfield  {author} {\bibinfo {author} {\bibfnamefont {P.}~\bibnamefont
  {Shor}},\ }\bibfield  {title} {\bibinfo {title} {Polynomial-time algorithms
  for prime factorization and discrete logarithms on a quantum computer},\
  }\href {https://doi.org/10.1137/S0097539795293172} {\bibfield  {journal}
  {\bibinfo  {journal} {SIAM J. Comput.}\ }\textbf {\bibinfo {volume} {26}},\
  \bibinfo {pages} {1484} (\bibinfo {year} {1997})}\BibitemShut {NoStop}%
\bibitem [{\citenamefont {Giovannetti}\ \emph {et~al.}(2004)\citenamefont
  {Giovannetti}, \citenamefont {Guha}, \citenamefont {Lloyd}, \citenamefont
  {Maccone}, \citenamefont {Shapiro},\ and\ \citenamefont
  {Yuen}}]{Giovannetti2004}%
  \BibitemOpen
  \bibfield  {author} {\bibinfo {author} {\bibfnamefont {V.}~\bibnamefont
  {Giovannetti}}, \bibinfo {author} {\bibfnamefont {S.}~\bibnamefont {Guha}},
  \bibinfo {author} {\bibfnamefont {S.}~\bibnamefont {Lloyd}}, \bibinfo
  {author} {\bibfnamefont {L.}~\bibnamefont {Maccone}}, \bibinfo {author}
  {\bibfnamefont {J.~H.}\ \bibnamefont {Shapiro}},\ and\ \bibinfo {author}
  {\bibfnamefont {H.~P.}\ \bibnamefont {Yuen}},\ }\bibfield  {title} {\bibinfo
  {title} {Classical capacity of the lossy bosonic channel: The exact
  solution},\ }\href {https://doi.org/10.1103/PhysRevLett.92.027902} {\bibfield
   {journal} {\bibinfo  {journal} {Phys. Rev. Lett.}\ }\textbf {\bibinfo
  {volume} {92}},\ \bibinfo {pages} {027902} (\bibinfo {year}
  {2004})}\BibitemShut {NoStop}%
\bibitem [{\citenamefont {Bennett}\ \emph {et~al.}(2002)\citenamefont
  {Bennett}, \citenamefont {Shor}, \citenamefont {Smolin},\ and\ \citenamefont
  {Thapliyal}}]{Bennett2002}%
  \BibitemOpen
  \bibfield  {author} {\bibinfo {author} {\bibfnamefont {C.}~\bibnamefont
  {Bennett}}, \bibinfo {author} {\bibfnamefont {P.}~\bibnamefont {Shor}},
  \bibinfo {author} {\bibfnamefont {J.}~\bibnamefont {Smolin}},\ and\ \bibinfo
  {author} {\bibfnamefont {A.}~\bibnamefont {Thapliyal}},\ }\bibfield  {title}
  {\bibinfo {title} {Entanglement-assisted capacity of a quantum channel and
  the reverse shannon theorem},\ }\href
  {https://doi.org/10.1109/TIT.2002.802612} {\bibfield  {journal} {\bibinfo
  {journal} {IEEE Trans. Inf. Theory}\ }\textbf {\bibinfo {volume} {48}},\
  \bibinfo {pages} {2637} (\bibinfo {year} {2002})}\BibitemShut {NoStop}%
\bibitem [{\citenamefont {Gisin}\ and\ \citenamefont
  {Thew}(2007)}]{gisin2007quantum}%
  \BibitemOpen
  \bibfield  {author} {\bibinfo {author} {\bibfnamefont {N.}~\bibnamefont
  {Gisin}}\ and\ \bibinfo {author} {\bibfnamefont {R.}~\bibnamefont {Thew}},\
  }\bibfield  {title} {\bibinfo {title} {Quantum communication},\ }\href@noop
  {} {\bibfield  {journal} {\bibinfo  {journal} {Nat. Photonics}\ }\textbf
  {\bibinfo {volume} {1}},\ \bibinfo {pages} {165} (\bibinfo {year}
  {2007})}\BibitemShut {NoStop}%
\bibitem [{\citenamefont {Lynch}(1996)}]{lynch1996distributed}%
  \BibitemOpen
  \bibfield  {author} {\bibinfo {author} {\bibfnamefont {N.~A.}\ \bibnamefont
  {Lynch}},\ }\href@noop {} {\emph {\bibinfo {title} {Distributed
  algorithms}}}\ (\bibinfo  {publisher} {Elsevier},\ \bibinfo {year}
  {1996})\BibitemShut {NoStop}%
\bibitem [{\citenamefont {Andrews}(2000)}]{andrews2000foundations}%
  \BibitemOpen
  \bibfield  {author} {\bibinfo {author} {\bibfnamefont {G.~R.}\ \bibnamefont
  {Andrews}},\ }\href@noop {} {\emph {\bibinfo {title} {Foundations of
  multithreaded, parallel, and distributed programming}}},\ Vol.~\bibinfo
  {volume} {11}\ (\bibinfo  {publisher} {Addison-Wesley Reading},\ \bibinfo
  {year} {2000})\BibitemShut {NoStop}%
\bibitem [{\citenamefont {Kimble}(2008)}]{kimble2008quantum}%
  \BibitemOpen
  \bibfield  {author} {\bibinfo {author} {\bibfnamefont {H.~J.}\ \bibnamefont
  {Kimble}},\ }\bibfield  {title} {\bibinfo {title} {The quantum internet},\
  }\href@noop {} {\bibfield  {journal} {\bibinfo  {journal} {Nature}\ }\textbf
  {\bibinfo {volume} {453}},\ \bibinfo {pages} {1023} (\bibinfo {year}
  {2008})}\BibitemShut {NoStop}%
\bibitem [{\citenamefont {Biamonte}\ \emph {et~al.}(2019)\citenamefont
  {Biamonte}, \citenamefont {Faccin},\ and\ \citenamefont
  {De~Domenico}}]{biamonte2019complex}%
  \BibitemOpen
  \bibfield  {author} {\bibinfo {author} {\bibfnamefont {J.}~\bibnamefont
  {Biamonte}}, \bibinfo {author} {\bibfnamefont {M.}~\bibnamefont {Faccin}},\
  and\ \bibinfo {author} {\bibfnamefont {M.}~\bibnamefont {De~Domenico}},\
  }\bibfield  {title} {\bibinfo {title} {Complex networks from classical to
  quantum},\ }\href@noop {} {\bibfield  {journal} {\bibinfo  {journal} {Commun.
  Phys.}\ }\textbf {\bibinfo {volume} {2}},\ \bibinfo {pages} {53} (\bibinfo
  {year} {2019})}\BibitemShut {NoStop}%
\bibitem [{\citenamefont {Wehner}\ \emph {et~al.}(2018)\citenamefont {Wehner},
  \citenamefont {Elkouss},\ and\ \citenamefont {Hanson}}]{wehner2018quantum}%
  \BibitemOpen
  \bibfield  {author} {\bibinfo {author} {\bibfnamefont {S.}~\bibnamefont
  {Wehner}}, \bibinfo {author} {\bibfnamefont {D.}~\bibnamefont {Elkouss}},\
  and\ \bibinfo {author} {\bibfnamefont {R.}~\bibnamefont {Hanson}},\
  }\bibfield  {title} {\bibinfo {title} {Quantum internet: A vision for the
  road ahead},\ }\href@noop {} {\bibfield  {journal} {\bibinfo  {journal}
  {Science}\ }\textbf {\bibinfo {volume} {362}} (\bibinfo {year}
  {2018})}\BibitemShut {NoStop}%
\bibitem [{\citenamefont {Kozlowski}\ and\ \citenamefont
  {Wehner}(2019)}]{kozlowski2019towards}%
  \BibitemOpen
  \bibfield  {author} {\bibinfo {author} {\bibfnamefont {W.}~\bibnamefont
  {Kozlowski}}\ and\ \bibinfo {author} {\bibfnamefont {S.}~\bibnamefont
  {Wehner}},\ }\bibfield  {title} {\bibinfo {title} {Towards large-scale
  quantum networks},\ }in\ \href@noop {} {\emph {\bibinfo {booktitle}
  {Proceedings of the Sixth Annual ACM International Conference on Nanoscale
  Computing and Communication}}}\ (\bibinfo {year} {2019})\ pp.\ \bibinfo
  {pages} {1--7}\BibitemShut {NoStop}%
\bibitem [{\citenamefont {Miguel-Ramiro}\ \emph {et~al.}(2020)\citenamefont
  {Miguel-Ramiro}, \citenamefont {Pirker},\ and\ \citenamefont
  {D{\"u}r}}]{miguel2020genuine}%
  \BibitemOpen
  \bibfield  {author} {\bibinfo {author} {\bibfnamefont {J.}~\bibnamefont
  {Miguel-Ramiro}}, \bibinfo {author} {\bibfnamefont {A.}~\bibnamefont
  {Pirker}},\ and\ \bibinfo {author} {\bibfnamefont {W.}~\bibnamefont
  {D{\"u}r}},\ }\bibfield  {title} {\bibinfo {title} {Genuine quantum networks:
  superposed tasks and addressing},\ }\href@noop {} {\bibfield  {journal}
  {\bibinfo  {journal} {arXiv:2005.00020}\ } (\bibinfo {year}
  {2020})}\BibitemShut {NoStop}%
\bibitem [{\citenamefont {Briegel}\ \emph {et~al.}(1998)\citenamefont
  {Briegel}, \citenamefont {D{\"u}r}, \citenamefont {Cirac},\ and\
  \citenamefont {Zoller}}]{briegel1998quantum}%
  \BibitemOpen
  \bibfield  {author} {\bibinfo {author} {\bibfnamefont {H.-J.}\ \bibnamefont
  {Briegel}}, \bibinfo {author} {\bibfnamefont {W.}~\bibnamefont {D{\"u}r}},
  \bibinfo {author} {\bibfnamefont {J.~I.}\ \bibnamefont {Cirac}},\ and\
  \bibinfo {author} {\bibfnamefont {P.}~\bibnamefont {Zoller}},\ }\bibfield
  {title} {\bibinfo {title} {Quantum repeaters: the role of imperfect local
  operations in quantum communication},\ }\href@noop {} {\bibfield  {journal}
  {\bibinfo  {journal} {Phys. Rev. Lett.}\ }\textbf {\bibinfo {volume} {81}},\
  \bibinfo {pages} {5932} (\bibinfo {year} {1998})}\BibitemShut {NoStop}%
\bibitem [{\citenamefont {Jiang}\ \emph {et~al.}(2009)\citenamefont {Jiang},
  \citenamefont {Taylor}, \citenamefont {Nemoto}, \citenamefont {Munro},
  \citenamefont {Van~Meter},\ and\ \citenamefont {Lukin}}]{jiang2009quantum}%
  \BibitemOpen
  \bibfield  {author} {\bibinfo {author} {\bibfnamefont {L.}~\bibnamefont
  {Jiang}}, \bibinfo {author} {\bibfnamefont {J.~M.}\ \bibnamefont {Taylor}},
  \bibinfo {author} {\bibfnamefont {K.}~\bibnamefont {Nemoto}}, \bibinfo
  {author} {\bibfnamefont {W.~J.}\ \bibnamefont {Munro}}, \bibinfo {author}
  {\bibfnamefont {R.}~\bibnamefont {Van~Meter}},\ and\ \bibinfo {author}
  {\bibfnamefont {M.~D.}\ \bibnamefont {Lukin}},\ }\bibfield  {title} {\bibinfo
  {title} {Quantum repeater with encoding},\ }\href
  {https://doi.org/10.1103/PhysRevA.79.032325} {\bibfield  {journal} {\bibinfo
  {journal} {Phys. Rev. A}\ }\textbf {\bibinfo {volume} {79}},\ \bibinfo
  {pages} {032325} (\bibinfo {year} {2009})}\BibitemShut {NoStop}%
\bibitem [{\citenamefont {Sangouard}\ \emph {et~al.}(2011)\citenamefont
  {Sangouard}, \citenamefont {Simon}, \citenamefont {De~Riedmatten},\ and\
  \citenamefont {Gisin}}]{sangouard2011quantum}%
  \BibitemOpen
  \bibfield  {author} {\bibinfo {author} {\bibfnamefont {N.}~\bibnamefont
  {Sangouard}}, \bibinfo {author} {\bibfnamefont {C.}~\bibnamefont {Simon}},
  \bibinfo {author} {\bibfnamefont {H.}~\bibnamefont {De~Riedmatten}},\ and\
  \bibinfo {author} {\bibfnamefont {N.}~\bibnamefont {Gisin}},\ }\bibfield
  {title} {\bibinfo {title} {Quantum repeaters based on atomic ensembles and
  linear optics},\ }\href@noop {} {\bibfield  {journal} {\bibinfo  {journal}
  {Rev. Mod. Phys.}\ }\textbf {\bibinfo {volume} {83}},\ \bibinfo {pages} {33}
  (\bibinfo {year} {2011})}\BibitemShut {NoStop}%
\bibitem [{\citenamefont {Munro}\ \emph {et~al.}(2015)\citenamefont {Munro},
  \citenamefont {Azuma}, \citenamefont {Tamaki},\ and\ \citenamefont
  {Nemoto}}]{munro2015inside}%
  \BibitemOpen
  \bibfield  {author} {\bibinfo {author} {\bibfnamefont {W.~J.}\ \bibnamefont
  {Munro}}, \bibinfo {author} {\bibfnamefont {K.}~\bibnamefont {Azuma}},
  \bibinfo {author} {\bibfnamefont {K.}~\bibnamefont {Tamaki}},\ and\ \bibinfo
  {author} {\bibfnamefont {K.}~\bibnamefont {Nemoto}},\ }\bibfield  {title}
  {\bibinfo {title} {Inside quantum repeaters},\ }\href@noop {} {\bibfield
  {journal} {\bibinfo  {journal} {IEEE Journal of Selected Topics in Quantum
  Electron.}\ }\textbf {\bibinfo {volume} {21}},\ \bibinfo {pages} {78}
  (\bibinfo {year} {2015})}\BibitemShut {NoStop}%
\bibitem [{\citenamefont {Muralidharan}\ \emph {et~al.}(2016)\citenamefont
  {Muralidharan}, \citenamefont {Li}, \citenamefont {Kim}, \citenamefont
  {L{\"u}tkenhaus}, \citenamefont {Lukin},\ and\ \citenamefont
  {Jiang}}]{muralidharan2016optimal}%
  \BibitemOpen
  \bibfield  {author} {\bibinfo {author} {\bibfnamefont {S.}~\bibnamefont
  {Muralidharan}}, \bibinfo {author} {\bibfnamefont {L.}~\bibnamefont {Li}},
  \bibinfo {author} {\bibfnamefont {J.}~\bibnamefont {Kim}}, \bibinfo {author}
  {\bibfnamefont {N.}~\bibnamefont {L{\"u}tkenhaus}}, \bibinfo {author}
  {\bibfnamefont {M.~D.}\ \bibnamefont {Lukin}},\ and\ \bibinfo {author}
  {\bibfnamefont {L.}~\bibnamefont {Jiang}},\ }\bibfield  {title} {\bibinfo
  {title} {Optimal architectures for long distance quantum communication},\
  }\href@noop {} {\bibfield  {journal} {\bibinfo  {journal} {Sci. Rep.}\
  }\textbf {\bibinfo {volume} {6}},\ \bibinfo {pages} {20463} (\bibinfo {year}
  {2016})}\BibitemShut {NoStop}%
\bibitem [{\citenamefont {Dias}\ and\ \citenamefont
  {Ralph}(2017)}]{dias2017quantum}%
  \BibitemOpen
  \bibfield  {author} {\bibinfo {author} {\bibfnamefont {J.}~\bibnamefont
  {Dias}}\ and\ \bibinfo {author} {\bibfnamefont {T.~C.}\ \bibnamefont
  {Ralph}},\ }\bibfield  {title} {\bibinfo {title} {Quantum repeaters using
  continuous-variable teleportation},\ }\href@noop {} {\bibfield  {journal}
  {\bibinfo  {journal} {Phys. Rev. A}\ }\textbf {\bibinfo {volume} {95}},\
  \bibinfo {pages} {022312} (\bibinfo {year} {2017})}\BibitemShut {NoStop}%
\bibitem [{\citenamefont {Muralidharan}\ \emph {et~al.}(2018)\citenamefont
  {Muralidharan}, \citenamefont {Zou}, \citenamefont {Li},\ and\ \citenamefont
  {Jiang}}]{jiang2018}%
  \BibitemOpen
  \bibfield  {author} {\bibinfo {author} {\bibfnamefont {S.}~\bibnamefont
  {Muralidharan}}, \bibinfo {author} {\bibfnamefont {C.-L.}\ \bibnamefont
  {Zou}}, \bibinfo {author} {\bibfnamefont {L.}~\bibnamefont {Li}},\ and\
  \bibinfo {author} {\bibfnamefont {L.}~\bibnamefont {Jiang}},\ }\bibfield
  {title} {\bibinfo {title} {One-way quantum repeaters with quantum
  reed-solomon codes},\ }\href {https://doi.org/10.1103/PhysRevA.97.052316}
  {\bibfield  {journal} {\bibinfo  {journal} {Phys. Rev. A}\ }\textbf {\bibinfo
  {volume} {97}},\ \bibinfo {pages} {052316} (\bibinfo {year}
  {2018})}\BibitemShut {NoStop}%
\bibitem [{\citenamefont {Furrer}\ and\ \citenamefont
  {Munro}(2018)}]{furrer2018repeaters}%
  \BibitemOpen
  \bibfield  {author} {\bibinfo {author} {\bibfnamefont {F.}~\bibnamefont
  {Furrer}}\ and\ \bibinfo {author} {\bibfnamefont {W.~J.}\ \bibnamefont
  {Munro}},\ }\bibfield  {title} {\bibinfo {title} {Repeaters for
  continuous-variable quantum communication},\ }\href@noop {} {\bibfield
  {journal} {\bibinfo  {journal} {Phys. Rev. A}\ }\textbf {\bibinfo {volume}
  {98}},\ \bibinfo {pages} {032335} (\bibinfo {year} {2018})}\BibitemShut
  {NoStop}%
\bibitem [{\citenamefont {Dias}\ \emph {et~al.}(2020)\citenamefont {Dias},
  \citenamefont {Hosseinidehaj},\ and\ \citenamefont
  {Ralph}}]{dias2020quantum}%
  \BibitemOpen
  \bibfield  {author} {\bibinfo {author} {\bibfnamefont {J.}~\bibnamefont
  {Dias}}, \bibinfo {author} {\bibfnamefont {N.}~\bibnamefont
  {Hosseinidehaj}},\ and\ \bibinfo {author} {\bibfnamefont {T.~C.}\
  \bibnamefont {Ralph}},\ }\bibfield  {title} {\bibinfo {title} {Quantum
  repeater for continuous variable entanglement distribution},\ }\href@noop {}
  {\bibfield  {journal} {\bibinfo  {journal} {arXiv:2004.06345}\ } (\bibinfo
  {year} {2020})}\BibitemShut {NoStop}%
\bibitem [{\citenamefont {Seshadreesan}\ \emph {et~al.}(2020)\citenamefont
  {Seshadreesan}, \citenamefont {Krovi},\ and\ \citenamefont
  {Guha}}]{Seshadreesan2020}%
  \BibitemOpen
  \bibfield  {author} {\bibinfo {author} {\bibfnamefont {K.~P.}\ \bibnamefont
  {Seshadreesan}}, \bibinfo {author} {\bibfnamefont {H.}~\bibnamefont
  {Krovi}},\ and\ \bibinfo {author} {\bibfnamefont {S.}~\bibnamefont {Guha}},\
  }\bibfield  {title} {\bibinfo {title} {Continuous-variable quantum repeater
  based on quantum scissors and mode multiplexing},\ }\href
  {https://doi.org/10.1103/PhysRevResearch.2.013310} {\bibfield  {journal}
  {\bibinfo  {journal} {Phys. Rev. Research}\ }\textbf {\bibinfo {volume}
  {2}},\ \bibinfo {pages} {013310} (\bibinfo {year} {2020})}\BibitemShut
  {NoStop}%
\bibitem [{\citenamefont {Goodenough}\ \emph {et~al.}(2020)\citenamefont
  {Goodenough}, \citenamefont {Elkouss},\ and\ \citenamefont
  {Wehner}}]{goodenough2020optimising}%
  \BibitemOpen
  \bibfield  {author} {\bibinfo {author} {\bibfnamefont {K.}~\bibnamefont
  {Goodenough}}, \bibinfo {author} {\bibfnamefont {D.}~\bibnamefont
  {Elkouss}},\ and\ \bibinfo {author} {\bibfnamefont {S.}~\bibnamefont
  {Wehner}},\ }\bibfield  {title} {\bibinfo {title} {Optimising repeater
  schemes for the quantum internet},\ }\href@noop {} {\bibfield  {journal}
  {\bibinfo  {journal} {arXiv:2006.12221}\ } (\bibinfo {year}
  {2020})}\BibitemShut {NoStop}%
\bibitem [{\citenamefont {Ac{\'\i}n}\ \emph {et~al.}(2007)\citenamefont
  {Ac{\'\i}n}, \citenamefont {Cirac},\ and\ \citenamefont
  {Lewenstein}}]{acin2007entanglement}%
  \BibitemOpen
  \bibfield  {author} {\bibinfo {author} {\bibfnamefont {A.}~\bibnamefont
  {Ac{\'\i}n}}, \bibinfo {author} {\bibfnamefont {J.~I.}\ \bibnamefont
  {Cirac}},\ and\ \bibinfo {author} {\bibfnamefont {M.}~\bibnamefont
  {Lewenstein}},\ }\bibfield  {title} {\bibinfo {title} {Entanglement
  percolation in quantum networks},\ }\href@noop {} {\bibfield  {journal}
  {\bibinfo  {journal} {Nat. Phys.}\ }\textbf {\bibinfo {volume} {3}},\
  \bibinfo {pages} {256} (\bibinfo {year} {2007})}\BibitemShut {NoStop}%
\bibitem [{\citenamefont {Brito}\ \emph {et~al.}(2020)\citenamefont {Brito},
  \citenamefont {Canabarro}, \citenamefont {Chaves},\ and\ \citenamefont
  {Cavalcanti}}]{brito2020statistical}%
  \BibitemOpen
  \bibfield  {author} {\bibinfo {author} {\bibfnamefont {S.}~\bibnamefont
  {Brito}}, \bibinfo {author} {\bibfnamefont {A.}~\bibnamefont {Canabarro}},
  \bibinfo {author} {\bibfnamefont {R.}~\bibnamefont {Chaves}},\ and\ \bibinfo
  {author} {\bibfnamefont {D.}~\bibnamefont {Cavalcanti}},\ }\bibfield  {title}
  {\bibinfo {title} {Statistical properties of the quantum internet},\
  }\href@noop {} {\bibfield  {journal} {\bibinfo  {journal} {Phys. Rev. Lett.}\
  }\textbf {\bibinfo {volume} {124}},\ \bibinfo {pages} {210501} (\bibinfo
  {year} {2020})}\BibitemShut {NoStop}%
\bibitem [{\citenamefont {Calderbank}\ and\ \citenamefont
  {Shor}(1996)}]{calderbank1996}%
  \BibitemOpen
  \bibfield  {author} {\bibinfo {author} {\bibfnamefont {A.~R.}\ \bibnamefont
  {Calderbank}}\ and\ \bibinfo {author} {\bibfnamefont {P.~W.}\ \bibnamefont
  {Shor}},\ }\bibfield  {title} {\bibinfo {title} {Good quantum
  error-correcting codes exist},\ }\href
  {https://doi.org/10.1103/PhysRevA.54.1098} {\bibfield  {journal} {\bibinfo
  {journal} {Phys. Rev. A}\ }\textbf {\bibinfo {volume} {54}},\ \bibinfo
  {pages} {1098} (\bibinfo {year} {1996})}\BibitemShut {NoStop}%
\bibitem [{\citenamefont {Caves}(1981)}]{caves1981quantum}%
  \BibitemOpen
  \bibfield  {author} {\bibinfo {author} {\bibfnamefont {C.~M.}\ \bibnamefont
  {Caves}},\ }\bibfield  {title} {\bibinfo {title} {Quantum-mechanical noise in
  an interferometer},\ }\href@noop {} {\bibfield  {journal} {\bibinfo
  {journal} {Phys. Rev. D}\ }\textbf {\bibinfo {volume} {23}},\ \bibinfo
  {pages} {1693} (\bibinfo {year} {1981})}\BibitemShut {NoStop}%
\bibitem [{\citenamefont {Gottesman}\ \emph {et~al.}(2001)\citenamefont
  {Gottesman}, \citenamefont {Kitaev},\ and\ \citenamefont
  {Preskill}}]{gottesman2001encoding}%
  \BibitemOpen
  \bibfield  {author} {\bibinfo {author} {\bibfnamefont {D.}~\bibnamefont
  {Gottesman}}, \bibinfo {author} {\bibfnamefont {A.}~\bibnamefont {Kitaev}},\
  and\ \bibinfo {author} {\bibfnamefont {J.}~\bibnamefont {Preskill}},\
  }\bibfield  {title} {\bibinfo {title} {Encoding a qubit in an oscillator},\
  }\href@noop {} {\bibfield  {journal} {\bibinfo  {journal} {Phys. Rev. A}\
  }\textbf {\bibinfo {volume} {64}},\ \bibinfo {pages} {012310} (\bibinfo
  {year} {2001})}\BibitemShut {NoStop}%
\bibitem{coutinho2021} 
B. C. Coutinho, W. J. Munro, K. Nemoto and Y. Omar, Robustness of Noisy Quantum Networks, arXiv:2103.03266 (2021)
\bibitem [{\citenamefont {Rabbie}\ \emph {et~al.}(2020)\citenamefont {Rabbie},
  \citenamefont {Chakraborty}, \citenamefont {Avis},\ and\ \citenamefont
  {Wehner}}]{rabbie2020designing}%
  \BibitemOpen
  \bibfield  {author} {\bibinfo {author} {\bibfnamefont {J.}~\bibnamefont
  {Rabbie}}, \bibinfo {author} {\bibfnamefont {K.}~\bibnamefont {Chakraborty}},
  \bibinfo {author} {\bibfnamefont {G.}~\bibnamefont {Avis}},\ and\ \bibinfo
  {author} {\bibfnamefont {S.}~\bibnamefont {Wehner}},\ }\bibfield  {title}
  {\bibinfo {title} {Designing quantum networks using preexisting
  infrastructure},\ }\href@noop {} {\bibfield  {journal} {\bibinfo  {journal}
  {arXiv:2005.14715}\ } (\bibinfo {year} {2020})}\BibitemShut {NoStop}%
\bibitem{pirandola2009}
S. Pirandola, R. Garc\'{i}a-Patr\'{o}n, S. L. Braunstein, and S. Lloyd, Direct and reverse secret-key capacities of aquantum channel, Phys. Rev. Lett.102, 050503 (2009).
\bibitem [{\citenamefont {Takeoka}\ \emph {et~al.}(2014)\citenamefont
  {Takeoka}, \citenamefont {Guha},\ and\ \citenamefont
  {Wilde}}]{takeoka2014fundamental}%
  \BibitemOpen
  \bibfield  {author} {\bibinfo {author} {\bibfnamefont {M.}~\bibnamefont
  {Takeoka}}, \bibinfo {author} {\bibfnamefont {S.}~\bibnamefont {Guha}},\ and\
  \bibinfo {author} {\bibfnamefont {M.~M.}\ \bibnamefont {Wilde}},\ }\bibfield
  {title} {\bibinfo {title} {Fundamental rate-loss tradeoff for optical quantum
  key distribution},\ }\href@noop {} {\bibfield  {journal} {\bibinfo  {journal}
  {Nat. Commun.}\ }\textbf {\bibinfo {volume} {5}},\ \bibinfo {pages} {5235}
  (\bibinfo {year} {2014})}\BibitemShut {NoStop}%
\bibitem [{\citenamefont {Pirandola}\ \emph {et~al.}(2017)\citenamefont
  {Pirandola}, \citenamefont {Laurenza}, \citenamefont {Ottaviani},\ and\
  \citenamefont {Banchi}}]{pirandola2017fundamental}%
  \BibitemOpen
  \bibfield  {author} {\bibinfo {author} {\bibfnamefont {S.}~\bibnamefont
  {Pirandola}}, \bibinfo {author} {\bibfnamefont {R.}~\bibnamefont {Laurenza}},
  \bibinfo {author} {\bibfnamefont {C.}~\bibnamefont {Ottaviani}},\ and\
  \bibinfo {author} {\bibfnamefont {L.}~\bibnamefont {Banchi}},\ }\bibfield
  {title} {\bibinfo {title} {Fundamental limits of repeaterless quantum
  communications},\ }\href@noop {} {\bibfield  {journal} {\bibinfo  {journal}
  {Nat. Commun.}\ }\textbf {\bibinfo {volume} {8}},\ \bibinfo {pages} {15043}
  (\bibinfo {year} {2017})}\BibitemShut {NoStop}%
\bibitem [{\citenamefont {Pirandola}(2019)}]{pirandola2019end}%
  \BibitemOpen
  \bibfield  {author} {\bibinfo {author} {\bibfnamefont {S.}~\bibnamefont
  {Pirandola}},\ }\bibfield  {title} {\bibinfo {title} {End-to-end capacities
  of a quantum communication network},\ }\href@noop {} {\bibfield  {journal}
  {\bibinfo  {journal} {Commun. Phys}\ }\textbf {\bibinfo {volume} {2}},\
  \bibinfo {pages} {51} (\bibinfo {year} {2019})};\ See also arXiv:1601.00966 (2016)\BibitemShut {NoStop}%
\QZ{
\bibitem [{\citenamefont {Azuma}\ \emph {et~al.}(2016)\citenamefont {Azuma},
  \citenamefont {Mizutani},\ and\ \citenamefont {Lo}}]{azuma2016fundamental}%
  \BibitemOpen
  \bibfield  {author} {\bibinfo {author} {\bibfnamefont {K.}~\bibnamefont
  {Azuma}}, \bibinfo {author} {\bibfnamefont {A.}~\bibnamefont {Mizutani}},\
  and\ \bibinfo {author} {\bibfnamefont {H.-K.}\ \bibnamefont {Lo}},\
  }\bibfield  {title} {\bibinfo {title} {Fundamental rate-loss trade-off for
  the quantum internet},\ }\href@noop {} {\bibfield  {journal} {\bibinfo
  {journal} {Nat. Commun.}\ }\textbf {\bibinfo {volume} {7}},\ \bibinfo {pages}
  { 13523} (\bibinfo {year} {2016})}\BibitemShut {NoStop}%
  }
\bibitem{patil2020} 
\QZ{
A. Patil, M. Pant, D. Englund, D. Towsley, S. Guha, Entanglement generation in a quantum network at distance-independent rate, 	arXiv:2005.07247 (2020)
}
\bibitem [{\citenamefont {Waxman}(1988)}]{waxman1988routing}%
  \BibitemOpen
  \bibfield  {author} {\bibinfo {author} {\bibfnamefont {B.~M.}\ \bibnamefont
  {Waxman}},\ }\bibfield  {title} {\bibinfo {title} {Routing of multipoint
  connections},\ }\href@noop {} {\bibfield  {journal} {\bibinfo  {journal}
  {IEEE J. Sel. Areas Commun.}\ }\textbf {\bibinfo {volume} {6}},\ \bibinfo
  {pages} {1617} (\bibinfo {year} {1988})}\BibitemShut {NoStop}%
\bibitem [{\citenamefont {Lakhina}\ \emph {et~al.}(2003)\citenamefont
  {Lakhina}, \citenamefont {Byers}, \citenamefont {Crovella},\ and\
  \citenamefont {Matta}}]{lakhina2003geographic}%
  \BibitemOpen
  \bibfield  {author} {\bibinfo {author} {\bibfnamefont {A.}~\bibnamefont
  {Lakhina}}, \bibinfo {author} {\bibfnamefont {J.~W.}\ \bibnamefont {Byers}},
  \bibinfo {author} {\bibfnamefont {M.}~\bibnamefont {Crovella}},\ and\
  \bibinfo {author} {\bibfnamefont {I.}~\bibnamefont {Matta}},\ }\bibfield
  {title} {\bibinfo {title} {On the geographic location of internet
  resources},\ }\href@noop {} {\bibfield  {journal} {\bibinfo  {journal} {IEEE
  J. Sel. Areas Commun.}\ }\textbf {\bibinfo {volume} {21}},\ \bibinfo {pages}
  {934} (\bibinfo {year} {2003})}\BibitemShut {NoStop}%
\bibitem [{\citenamefont {Barab{\'a}si}\ and\ \citenamefont
  {Albert}(1999)}]{barabasi1999emergence}%
  \BibitemOpen
  \bibfield  {author} {\bibinfo {author} {\bibfnamefont {A.-L.}\ \bibnamefont
  {Barab{\'a}si}}\ and\ \bibinfo {author} {\bibfnamefont {R.}~\bibnamefont
  {Albert}},\ }\bibfield  {title} {\bibinfo {title} {Emergence of scaling in
  random networks},\ }\href@noop {} {\bibfield  {journal} {\bibinfo  {journal}
  {science}\ }\textbf {\bibinfo {volume} {286}},\ \bibinfo {pages} {509}
  (\bibinfo {year} {1999})}\BibitemShut {NoStop}%
\bibitem [{\citenamefont {Yook}\ \emph {et~al.}(2002)\citenamefont {Yook},
  \citenamefont {Jeong},\ and\ \citenamefont
  {Barab{\'a}si}}]{yook2002modeling}%
  \BibitemOpen
  \bibfield  {author} {\bibinfo {author} {\bibfnamefont {S.-H.}\ \bibnamefont
  {Yook}}, \bibinfo {author} {\bibfnamefont {H.}~\bibnamefont {Jeong}},\ and\
  \bibinfo {author} {\bibfnamefont {A.-L.}\ \bibnamefont {Barab{\'a}si}},\
  }\bibfield  {title} {\bibinfo {title} {Modeling the internet's large-scale
  topology},\ }\href@noop {} {\bibfield  {journal} {\bibinfo  {journal} {Proc.
  Natl. Acad. Sci.}\ }\textbf {\bibinfo {volume} {99}},\ \bibinfo {pages}
  {13382} (\bibinfo {year} {2002})}\BibitemShut {NoStop}%
\bibitem [{\citenamefont {Pastor-Satorras}\ and\ \citenamefont
  {Vespignani}(2007)}]{pastor2007evolution}%
  \BibitemOpen
  \bibfield  {author} {\bibinfo {author} {\bibfnamefont {R.}~\bibnamefont
  {Pastor-Satorras}}\ and\ \bibinfo {author} {\bibfnamefont {A.}~\bibnamefont
  {Vespignani}},\ }\href@noop {} {\emph {\bibinfo {title} {Evolution and
  structure of the Internet: A statistical physics approach}}}\ (\bibinfo
  {publisher} {Cambridge University Press},\ \bibinfo {year}
  {2007})\BibitemShut {NoStop}%
\bibitem [{\citenamefont {Zhang}\ and\ \citenamefont
  {Zhuang}(2020)}]{zhang2020entanglement}%
  \BibitemOpen
  \bibfield  {author} {\bibinfo {author} {\bibfnamefont {B.}~\bibnamefont
  {Zhang}}\ and\ \bibinfo {author} {\bibfnamefont {Q.}~\bibnamefont {Zhuang}},\
  }\bibfield  {title} {\bibinfo {title} {Entanglement formation in
  continuous-variable random quantum networks},\ }\href@noop {} {\bibfield
  {journal} {\bibinfo  {journal} {arXiv:2005.12934}\ } (\bibinfo {year}
  {2020})}\BibitemShut {NoStop}%
\bibitem [{\citenamefont {Mukherjee}(2000)}]{mukherjee2000wdm}%
  \BibitemOpen
\bibfield  {journal} {  }\bibfield  {author} {\bibinfo {author} {\bibfnamefont
  {B.}~\bibnamefont {Mukherjee}},\ }\bibfield  {title} {\bibinfo {title} {Wdm
  optical communication networks: progress and challenges},\ }\href@noop {}
  {\bibfield  {journal} {\bibinfo  {journal} {IEEE J. Sel. Areas Commun.}\
  }\textbf {\bibinfo {volume} {18}},\ \bibinfo {pages} {1810} (\bibinfo {year}
  {2000})}\BibitemShut {NoStop}%
\bibitem [{\citenamefont {Zukowski}\ \emph {et~al.}(1993)\citenamefont
  {Zukowski}, \citenamefont {Zeilinger}, \citenamefont {Horne},\ and\
  \citenamefont {Ekert}}]{zukowski1993event}%
  \BibitemOpen
  \bibfield  {author} {\bibinfo {author} {\bibfnamefont {M.}~\bibnamefont
  {Zukowski}}, \bibinfo {author} {\bibfnamefont {A.}~\bibnamefont {Zeilinger}},
  \bibinfo {author} {\bibfnamefont {M.~A.}\ \bibnamefont {Horne}},\ and\
  \bibinfo {author} {\bibfnamefont {A.~K.}\ \bibnamefont {Ekert}},\ }\bibfield
  {title} {\bibinfo {title} {" event-ready-detectors" bell experiment via
  entanglement swapping.},\ }\href@noop {} {\bibfield  {journal} {\bibinfo
  {journal} {Phys. Rev. Lett.}\ }\textbf {\bibinfo {volume} {71}} (\bibinfo
  {year} {1993})}\BibitemShut {NoStop}%
\bibitem{GiovannettiV2014}%
V. Giovannetti, R. Garc\'{i}a-Patr\'{o}n, N. J. Cerf and A. S. Holevo, Ultimate classical
communication rates of quantum optical channels, Nat
Photon {\bf 8}, 796 (2014).
\bibitem{poisson}
The degree distribution holds in Poisson form when nodes density is small.
\end{thebibliography}

\appendix

\section{Basic properties networks.}
\label{appA}
\begin{figure}[b]
    \centering
    \includegraphics[width = 0.5\textwidth]{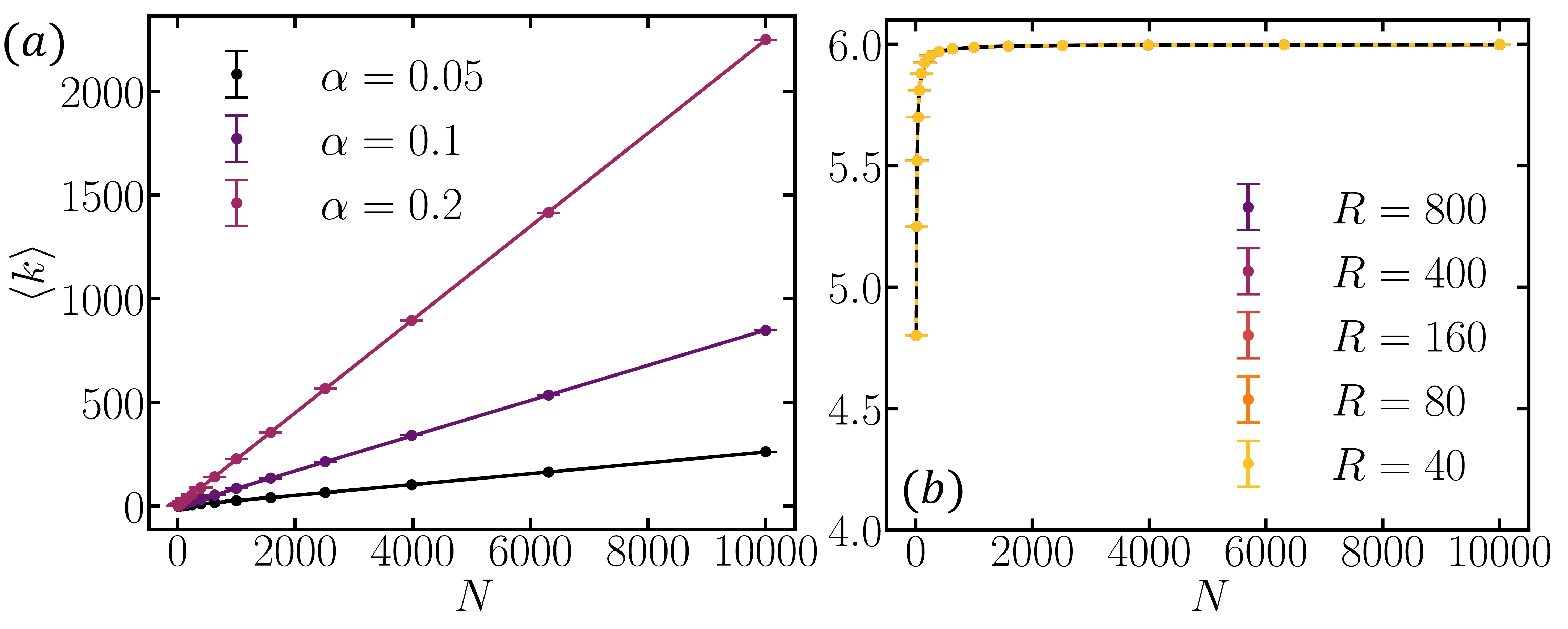}
    \caption{Average degree of (a) Waxman and (b) Yook models and its dependence on the number of nodes $N$. (a) Solid lines gives linear fitting results of $\braket{k}=A\rho$ where $\rho$ is the density of nodes. (b) Dashed lines show the theory curve $\braket{k}=(2N-1-m)m/N$.
    \label{fig:mean_degree}
    }
    \centering
    \includegraphics[width = 0.5\textwidth]{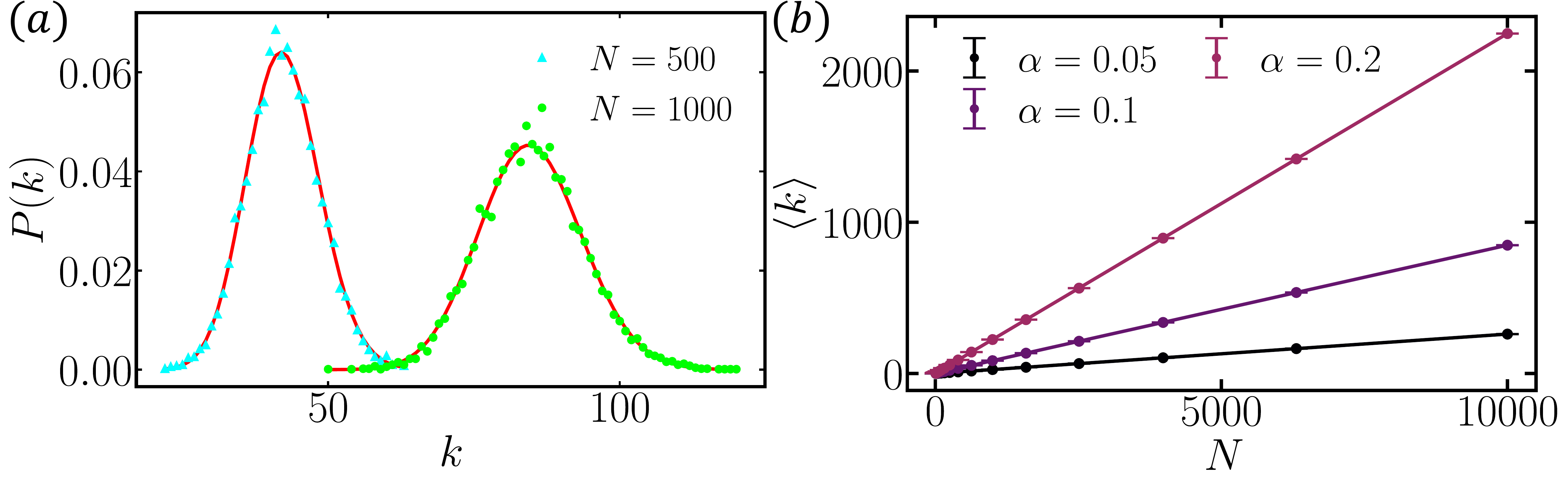}
    \caption{Degree distribution and mean degree of Erd\H{o}s R\'enyi model. (a) Degree distribution of Erd\H{o}s R\'enyi model with $\alpha=0.1$. The red ccurves represent the analytical expression for it. (b) Mean degree of Erd\H{o}s R\'enyi model with different $\alpha$.}
    \label{fig:ER_degree}
\end{figure}

As shown in Fig.~\ref{fig:mean_degree}, in the Waxman model, the average degree $\braket{k}$ of the nodes increases with the number of nodes $N$ linearly, at a rate depending on the scale $\alpha$; In the scale-free network, the average degree saturates to $\braket{k}=(2N-1-m)m/N\simeq2m$ as the number of nodes $N$ increases. Here $m$ is the number of edges brought by the addition of each single node.

We plot the degree distribution of Erd\H{o}s R\'enyi model and mean degree in Fig.~\ref{fig:ER_degree}.

\begin{figure*}
\centering
\includegraphics[width=0.9\textwidth]{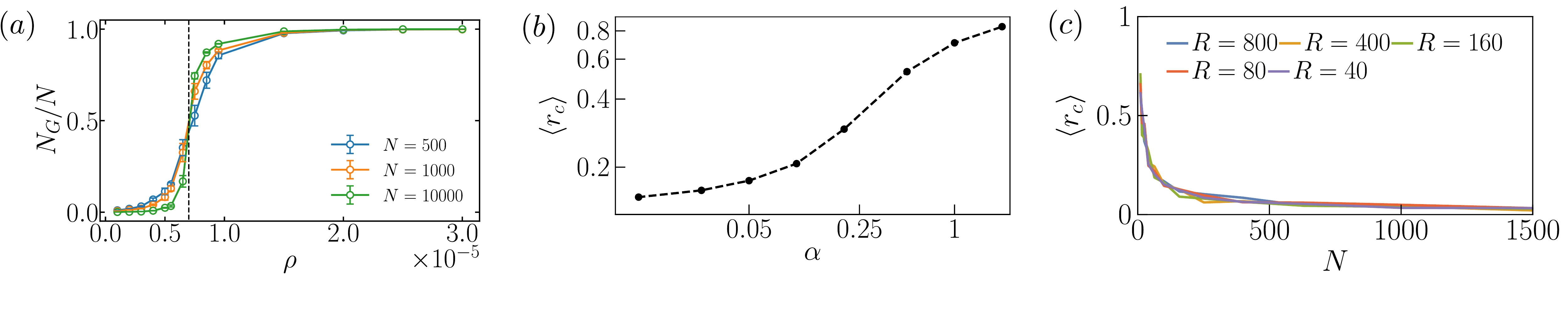}
\caption{
(a) The relative size $N_G/N$ of the largest component in the Waxman \acrshort{QN} model vs. the density of nodes, $\alpha, L$ are both determined by $N$ and density. To obtain the average, we sampled $10,10,5$ graphs for $N = 500, 1000, 10^4$ separately. The dashed vertical line at a density $\sim 7\times10^{-6}$ indicates the transition point.
(b) Clustering coefficients vs $\alpha$ for Waxman \acrshort{QN} model, in the large number of nodes $N\gg1$ limit.
(c) Clustering coefficients of the scale-free \acrshort{QN} model.
\label{fig:giant_rc_alpha}
}
\end{figure*}

The Waxman model has a giant component transition as the density of nodes $\rho$ increases. As shown in Fig.~\ref{fig:giant_rc_alpha}(a), the ratio of the size of the largest connected component $N_G$ over the total number of nodes $N$ increases from close to zero to unity abruptly at a density of $\rho_G\simeq 7\times 10^{-6}$. The transition becomes sharper as the number of nodes increase.

To understand the connectivity of the networks, we plot the clustering coefficient's dependence network parameters. For a single node, the single-node local clustering coefficient $r_c(\bm x) = t/[k(k-1)/2]$ identifies the existence of connections between its $k$ neighbors $\calN(\bm x)$. Here $t$ is number of triangles that is attached to the node $\bm x$. We can define the graph clustering coefficient $\braket{r_c}$ by averaging over all nodes. For Waxman networks, $\braket{r_c}$ converging to a constant dependent on $\alpha$ as the number of nodes increases, as shown in Fig.~\ref{fig:giant_rc_alpha}(b). While for the scale-free networks, $\braket{r_c}$ decays to zero as the number of nodes $N$ increases, as shown in Fig.~\ref{fig:giant_rc_alpha}(c).

\section{Additional data for the end-to-end capacity}
\label{appB}
\begin{figure*}
\centering
\includegraphics[width = 1\textwidth]{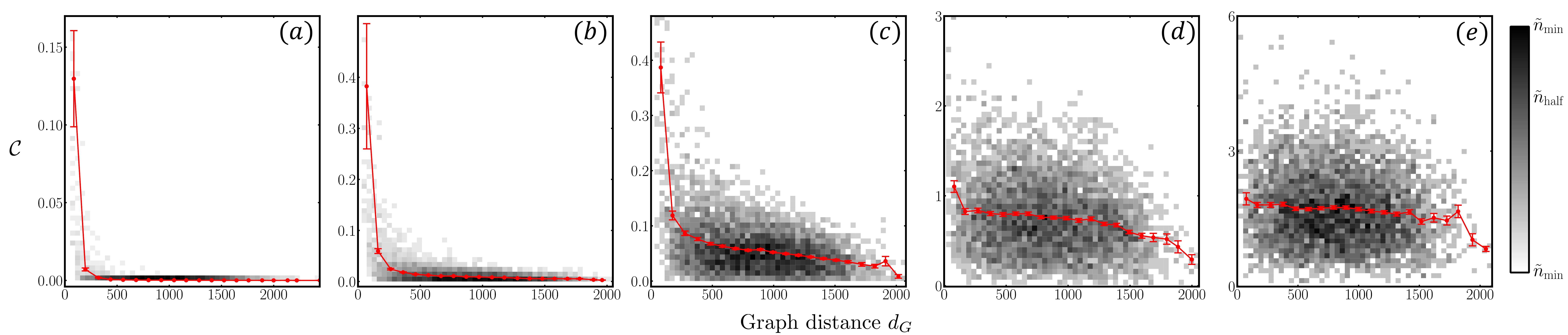}
\caption{More details on Fig.~\ref{fig:graphs_full}(b1)-(b5). We have used the same numbering of the subplots for consistency. The gray scale PDF represents the statistical distribution (plotted in nonlinear scale $\sqrt{\tilde{n}}$ for visualization) of end-to-end capacity over 5000 random pairs of end nodes (50 pairs from each of the 100 random \acrshort{QN}s). The red lines are the average end-to-end capacity in each of the distance window. We sort the 5000 samples according to the graph distances from small to large and divided them into 20 groups of 250 points accordingly. We take the average of the capacity and graph distance in each group and obtain a data point.
\label{fig:graphs_wax_dis}
}
\centering
\includegraphics[width = 0.5\textwidth]{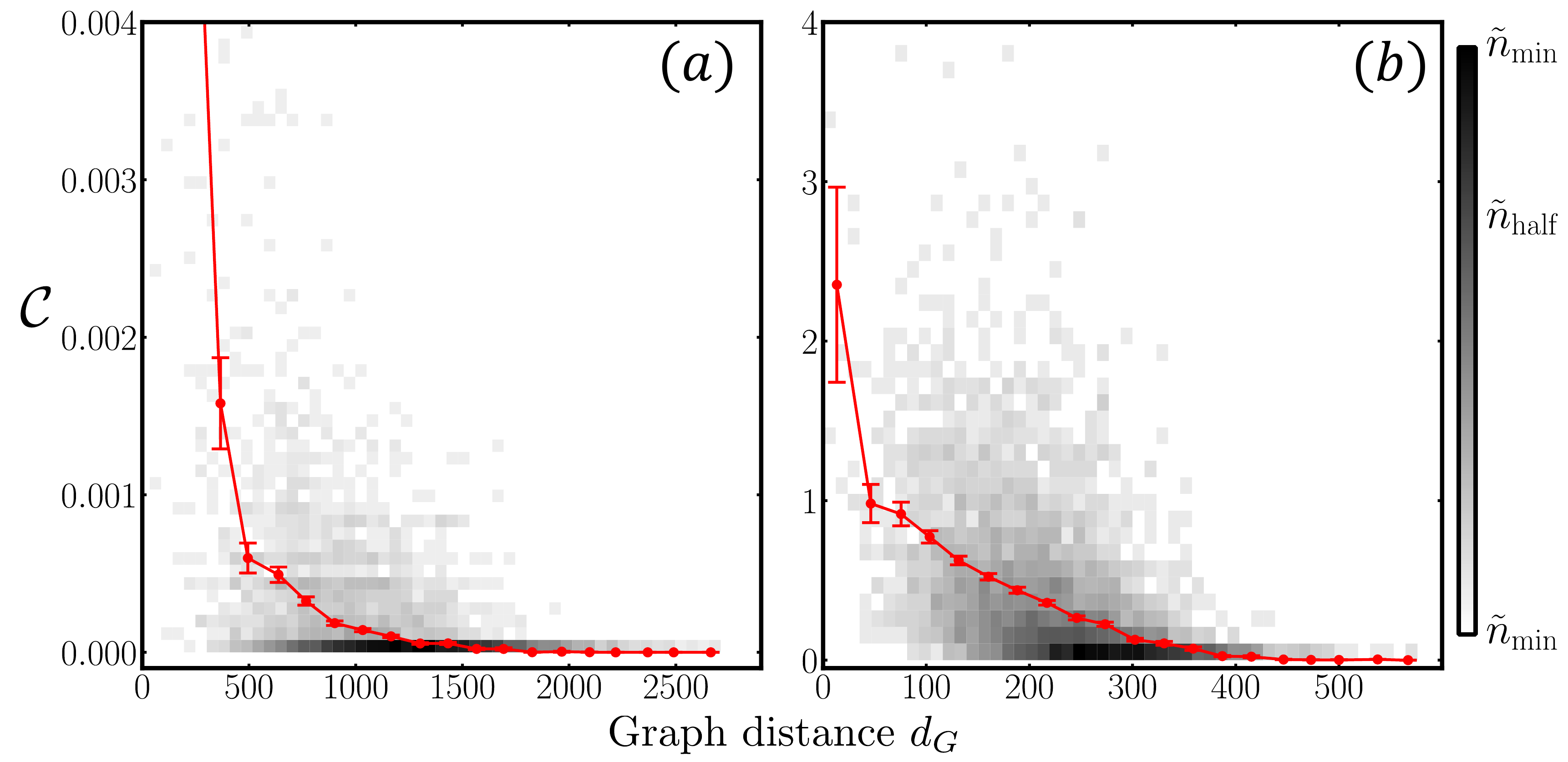}
\caption{More details on Fig.~\ref{fig:graphs_YOOK}(b1)(b2) in the main paper. We have used the same numbering of the subplots for consistency. The gray scale PDF represents the statistical distribution (plotted in nonlinear scale $\sqrt{\tilde{n}}$ for visualization) of end-to-end capacity over 5000 random pairs of end nodes (50 pairs from each of the 100 random \acrshort{QN}s). The red lines are the average end-to-end capacity in each of the distance window. We sort the 5000 samples according to the graph distances from small to large and divided them into 20 groups of 250 points accordingly. We take the average of the capacity and graph distance in each group and obtain a data point.
 \label{fig:graphs_YOOK_dis}
}
\end{figure*}

We provide additional data of the numerical calculations.
First, we show the distribution of the end-to-end capacity between random pairs of nodes in each ensemble of networks. Fig.~\ref{fig:graphs_wax_dis} shows the Waxman case, corresponding to Fig.~\ref{fig:graphs_full}(b1)-(b5); while Fig.~\ref{fig:graphs_YOOK_dis} shows the scale-free case, corresponding to Fig.~\ref{fig:graphs_YOOK}(b1)(b2) of the main paper. The average of the data utilized here gives the red curves in the corresponding plots of the main paper, which are also shown as red curves in these plots.

\begin{figure*}
    \centering
    \includegraphics[width = 1\textwidth]{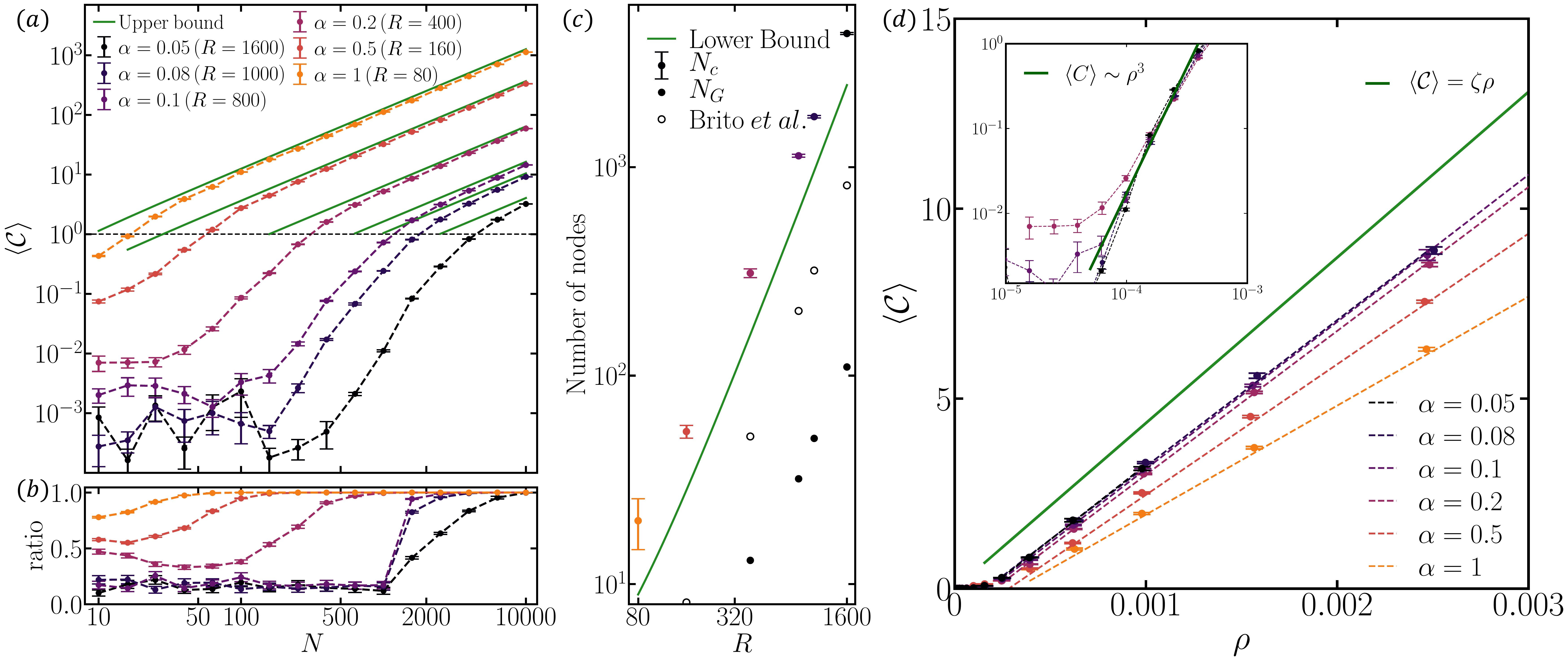}
    \caption{(a) Average end-to-end capacity $\braket{\calC}$ vs. number of nodes $N$ for various $\alpha$'s. The system size $R\simeq 80/\alpha$ km. We see a critical drop at small $N$ when $\alpha$ is not too large, indicated by the dashed lines going down to zero. The green solid lines gives the upper bounds in Eq.~\eqref{C_UB}.
    (b) The average of the ratio of end-node edges inside the minimum cut. It shares the same legend as in (a).
    (c) We plot the critical number of nodes vs $\alpha$. The green curve indicates a lower bound from solving $\braket{\calC\left(\bm x\right)}=1$ in Eq.~\eqref{C_UB}. In comparison, we plotted the results predicted from Ref.~\cite{brito2020statistical} (black open circles) and the critical number of nodes $N_G$ for the appearance of giant components.
    (d) Average end-to-end capacity $\braket{\calC}$ vs. density of nodes $\rho$ for various $\alpha$'s in a linear scale. The system size $R\simeq 80/\alpha$ km. The green line represents the asymptotic upper bounds $\braket{\calC} \simeq \zeta \rho$ and the dashed lines with same colors as dots shows the linear fitting in the range $\braket{\calC}>0.1$. The inset is the dependence of average end-to-end capacity with density in the range $\rho\in(10^{-5}, 10^{-3})$. The dark blue solid line presents a power-law relation as $\braket{\calC} \sim \rho^3$.
    }
    \label{fig:C_N}
\end{figure*}

\begin{figure}
    \centering
    \includegraphics[width = 0.5\textwidth]{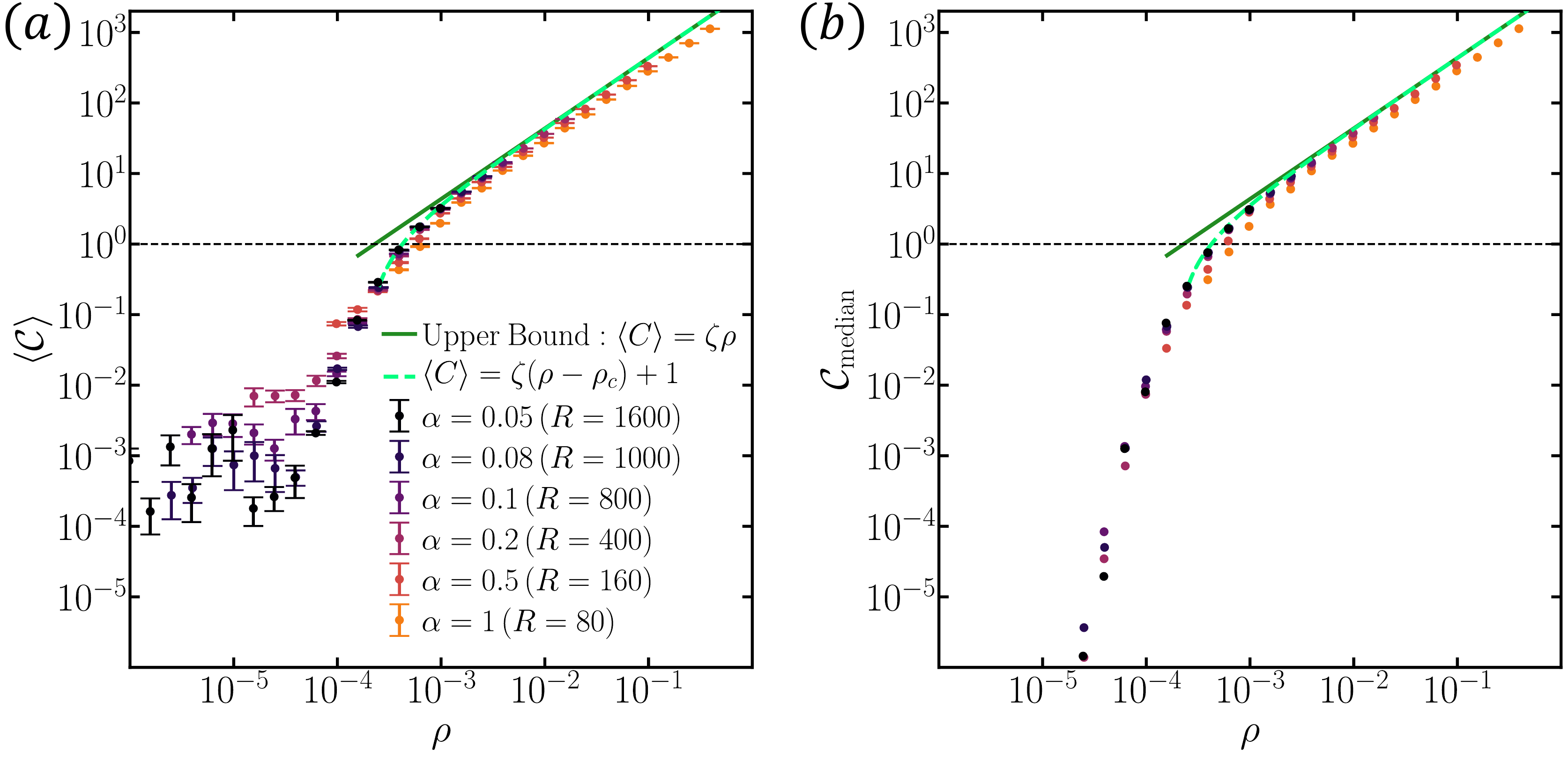}
    \caption{(a) Average end-to-end capacity $\braket{\calC}$ vs. density of nodes $\rho$ for various $\alpha$'s. The system size $R\simeq 80/\alpha$ km. We see a critical drop at small $N$ when $\alpha$ is not too large, indicated by the dashed lines going down to zero. The green solid lines gives the upper bounds in Eq.~\eqref{C_UB}.
    (b) The median end-to-end capacity $\braket{\calC}$ vs. density of nodes $\rho$ for various $\alpha$'s. It shares the same legend as in (a). Instead of a long tail, we see clear sharp drop around the transition point.
    }
    \label{fig:C_rho_median}
    \centering
    \includegraphics[width = 0.3\textwidth]{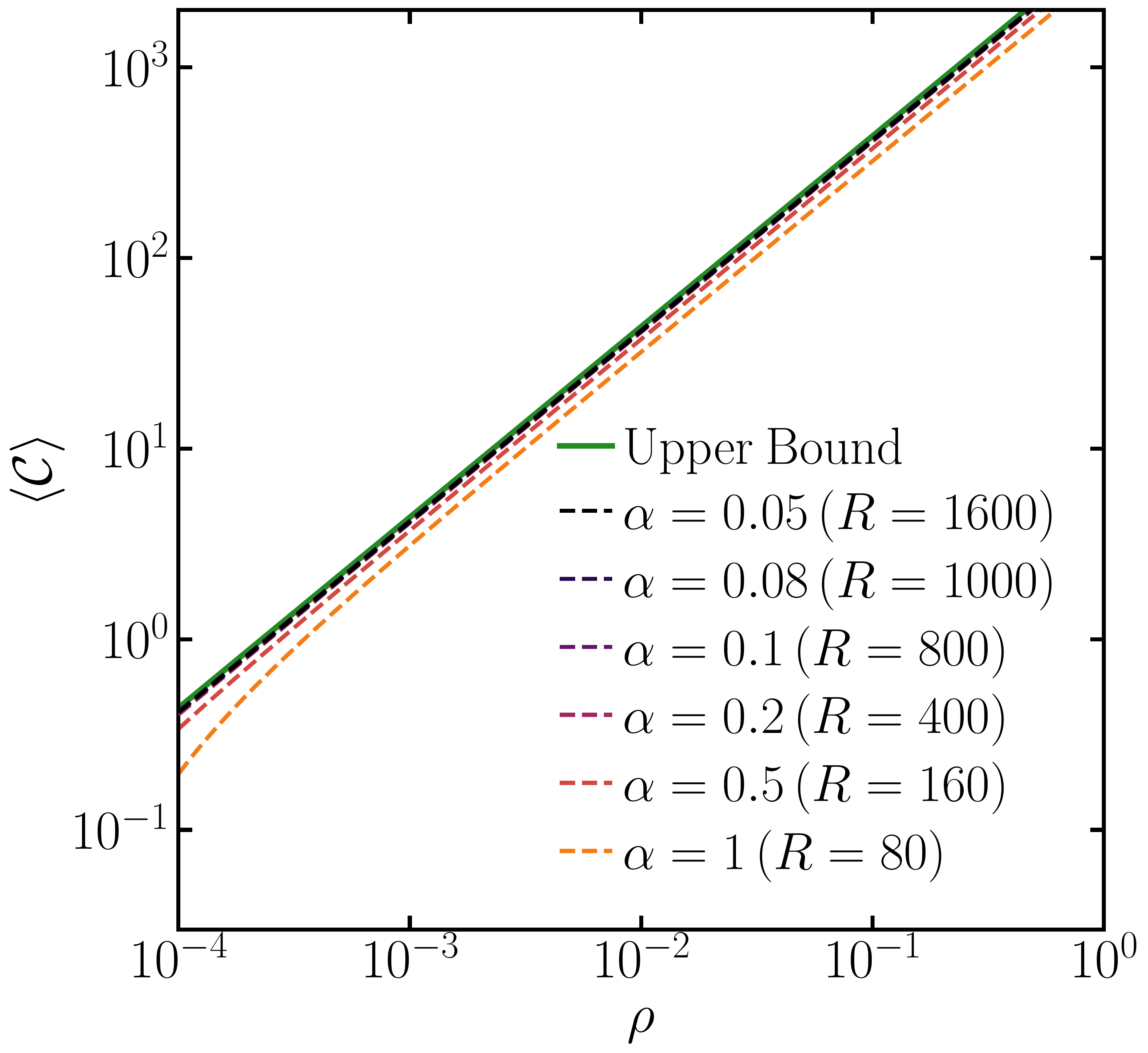}
    \caption{Comparison of the exact upper bound in Eq.~\eqref{C_UB} and its asymptotic limit $\braket{C}\simeq\zeta\rho$ for the Waxman model. We plot those upper bound by dashed lines with different $\alpha$ and asymptotic limit in orange line.}
    \label{fig:waxman_UB_compare}
\end{figure}

Next, we present an in-depth analyses of Fig.~\ref{fig:C_rho_alpha} in the main paper. Fig.~\ref{fig:C_N}(a) shows each curve of capacity vs. number of nodes for different scales individually, without collapsing everything in plotting with density. In the main paper, we do not show the long tails, as these tails are mainly due to rare cases of random pairs of nodes lying very close to each other. Indeed, if we plot the median instead of the mean, as shown in Fig.~\ref{fig:C_rho_median}, these long tails are not present and we see a clear sharp drop. To avoid burying the main take-away in such technical details, we do not present the entire data in the main paper. Here we also evaluated the exact upper bound from Eq.~\eqref{C_UB} for each curve, which converges to the asymptotic results shown in the main paper (see Fig.~\ref{fig:waxman_UB_compare} for details of the convergence). In Fig.~\ref{fig:C_N} (b), we calculate the critical number $N_c$ for $\braket{C}=1$, which is much larger than the giant component transition point $N_G$ or the results from Ref.~\cite{brito2020statistical}. 
We can also solve $\braket{\calC\left(\bm x\right)}=1$ in Eq.~\eqref{C_UB} to obtain a lower bound estimate on $N_c$, which works well when $R$ is large as shown in Fig.~\ref{fig:C_N}(c). In Fig.~\ref{fig:C_N}(d), we plot the capacity in linear scale to show the deviations between the actual capacities and the upper bounds in more detail. The major reason for the deviation at large $R$ and high density is due to the second inequality of Ineq.~\eqref{C_end_to_end_UB_further} of the main paper, which we also print here
\be 
\braket{\calC \left(\bm x, \bm x^\prime\right)}\le  \braket{\min\left\{\calC\left(\bm x\right),\calC\left(\bm x^\prime\right)\right\}}\le \braket{\calC\left(\bm x\right)},
\ee 
as interchanging the order of ensemble averaging and minimization is not tight.

\section{Derivation of the asymptotic results for Waxman model}
\label{appC}
Due to the independence between the edges between nodes, we have
\begin{widetext}
\begin{align}
&\braket{\calC\left(\bm x\right)}=\frac{(N-1)}{|\Omega_R|^2}\int_{\Omega_R} d^2\bm  x \int_{\Omega_R} d^2 \bm x^\prime   \Pi\left(\bm x, \bm x^\prime \right) \calC \left(E_{\bm x, \bm x^\prime}\right)
\label{C_UB}
\\ 
&=-\frac{(N-1)}{|\Omega_R|^2}\int_{\Omega_R} d^2\bm  x \int_{\Omega_R} d^2 \bm x^\prime  e^{-D(\bm x, \bm x^\prime)/\alpha L} \log_2\left(1-10^{-\gamma D(\bm x, \bm x^\prime)} \right)
\\
&=-\frac{(N-1)}{|\Omega_R|^2}\left[\int_{\Omega_R} d^2\bm  x 
\int_{\Omega_\infty} d^2 \bm x^\prime
e^{-D(\bm x, \bm x^\prime)/\alpha L} \log_2\left(1-10^{-\gamma D(\bm x, \bm x^\prime)} \right)
+O(R)
\right]
\\
&
=-\frac{(N-1)}{|\Omega_R|}
\int_{\Omega_\infty} d^2 \bm x^\prime
e^{-D(\bm x, \bm x^\prime)/\alpha L} \log_2\left(1-10^{-\gamma D(\bm x, \bm x^\prime)} \right)
+O(NR^{-3})
\\
&
=-\frac{(N-1)\pi}{2R^2}
\int_{0}^\infty  r \ dr\ 
e^{-r/\alpha L} \log_2\left(1-10^{-\gamma r} \right)
+O(NR^{-3})
\\
&
=-\frac{(N-1)\pi}{2R^2}
\int_{0}^\infty  r \ dr\ 
e^{-r/\alpha L} \log_2\left(1-10^{-\gamma r} \right)
+O(NR^{-3})
\\
&
=-2\pi\rho
\int_{0}^\infty  r \ dr\ 
e^{-r/\alpha L} \log_2\left(1-10^{-\gamma r} \right)
+O(NR^{-3})+O(R^{-2}).
\end{align}
\end{widetext}
Inputting $\alpha L=226$ and $\gamma=0.02$ we have the asymptotic expansion of 
\begin{align}
&\braket{\calC\left(\bm x\right)}=\zeta \rho,
\\
&\zeta=-2\pi\int_0^\infty dr \ r e^{-r/226}\log_2\left(1-10^{-0.02 r}\right)\simeq 4357.9.
\label{zeta_asym}
\end{align} 
In Fig.~\ref{fig:waxman_UB_compare}, we compare the asymptotic results with the exact numerical integration in Eq.~\eqref{C_UB}. A good convergence towards the asymptotic result is found with the increasing scale $R$.

\section{Derivation of node capacity in Erd\H{o}s R\'enyi models}
\label{appD}
\begin{equation}
\begin{split}
&\braket{\calC\left(\bm x\right)} =\frac{(N-1)}{|\Omega_R|^2}\int_{\Omega_R} d^2\bm  x \int_{\Omega_R} d^2 \bm x^\prime   \Pi\left(\bm x, \bm x^\prime \right) \calC_E \left(E_{\bm x, \bm x^\prime}\right)\\ 
&=-\frac{(N-1)p}{|\Omega_R|^2}\int_{\Omega_R} d^2\bm  x \int_{\Omega_R} d^2 \bm x^\prime \log_2\left(1-10^{-\gamma D(\bm x, \bm x^\prime)} \right)\\
&=-\frac{(N-1)p}{|\Omega_R|}
\int_{\Omega_\infty} d^2 \bm x^\prime
\log_2\left(1-10^{-\gamma D(\bm x, \bm x^\prime)} \right)
+O(NR^{-3})\\
&=-\frac{(N-1)p\pi}{2R^2}\int_{0}^\infty  r \ dr \log_2\left(1-10^{-\gamma r} \right)
+O(NR^{-3})\\
&=\zeta_{ER} p \rho +O(NR^{-3})+O(R^{-2}),
\end{split}
\end{equation}

\section{Derivation of the asymptotic results for scale-free model} 
\label{appE}
Considering the on average $2m$ neighbours as independent, the ensemble-averaged node capacity is 
\begin{align}
&\braket{\calC\left(\bm x\right)}=\frac{2m}{|\Omega_R|^2}\int_{\Omega_R} d^2\bm  x \int_{\Omega_R} d^2 \bm x^\prime   \braket{\Pi\left(\bm x, \bm x^\prime \right) \calC \left(E_{\bm x, \bm x^\prime}\right)}
\label{C_UB_YOOK}
\\
&=
\frac{2m}{ A}\int_{\Omega_R} d^2\bm  x \int_{\Omega_R} d^2 \bm x^\prime    \braket{\frac{D_g\left(\bm x^\prime\right)}{D\left(\bm x, \bm x^\prime\right)} \calC \left(E_{\bm x, \bm x^\prime}\right)}
\end{align}
where the normalization constant
\be 
A=\int_{\Omega_R} d^2\bm  x \int_{\Omega_R} d^2 \bm x^\prime    \braket{\frac{D_g\left(\bm x^\prime\right)}{D\left(\bm x, \bm x^\prime\right)} }.
\ee
The $\braket{\cdot}$ inside the integral now denotes average over the degree distribution of neighbours, conditioned on the neighbour being at $\bm x^\prime$.
We can approximate the distribution of the degree as independent of the distance to node $\bm x$, then $\braket{D_g\left(\bm x^\prime\right) f(\bm x,\bm x^\prime)}=\braket{D}f(\bm x,\bm x^\prime)$, where $\braket{D}$ is a constant and $f(\bm x,\bm x^\prime)$ is an arbitrary function of $\bm x, \bm x^\prime$. We can cancel out the constant and equivalently calculate
\begin{align}
&\braket{\calC\left(\bm x\right)}=
\frac{2m}{ A^\prime}\int_{\Omega_R} d^2\bm  x \int_{\Omega_R} d^2 \bm x^\prime    \braket{\frac{1}{D\left(\bm x, \bm x^\prime\right)} \calC \left(E_{\bm x, \bm x^\prime}\right)}
\\
&A^\prime=\int_{\Omega_R} d^2\bm  x \int_{\Omega_R} d^2 \bm x^\prime    \braket{\frac{1}{D\left(\bm x, \bm x^\prime\right)} }.
\end{align}
The above integral can be numerically calculated. It is clear that $\braket{\calC\left(\bm x\right)}$ does not grow with the number of nodes $N$, as $m$ is now a constant.

\newpage

\end{document}